\documentclass[fleqn,10pt]{wlscirep}
\usepackage[utf8]{inputenc}
\usepackage[T1]{fontenc}
\usepackage{subcaption}
\usepackage{svg}
\usepackage{lineno}
\usepackage{array}
\usepackage{placeins}
\usepackage{needspace}
\usepackage{listings}
\lstdefinestyle{jsonstyle}{
    basicstyle=\ttfamily\scriptsize,
    breaklines=true,
    breakatwhitespace=false,
    frame=single,
    rulecolor=\color{gray},
    backgroundcolor=\color{gray!5},
    columns=fullflexible,
    showstringspaces=false,
    xleftmargin=4pt,
    xrightmargin=4pt,
    aboveskip=4pt,
    belowskip=2pt,
}
\title{GazeXPErT: An Expert Eye-tracking Dataset for Interpretable and Explainable AI in Oncologic FDG-PET/CT Scans}

\author[1,$\dag$,*]{Joy T Wu}
\author[2,$\dag$,*]{Daniel Beckmann}
\author[1]{Sarah Miller}
\author[1]{Alexander Lee}
\author[1]{Elizabeth Theng}
\author[1]{Stephan Altmayer}
\author[1]{Ken Chang}
\author[1]{David Kersting}
\author[1]{Tomoaki Otani}
\author[1]{Brittany Z Dashevsky}
\author[1]{Hye Lim Park}
\author[1]{Matteo Novello}
\author[1]{Kip Guja}
\author[1]{Curtis Langlotz}
\author[4]{Ismini Lourentzou}
\author[3]{Daniel Gruhl}
\author[2,$\dag$$\dag$]{Benjamin Risse}
\author[1,$\dag$$\dag$]{Guido A Davidzon}

\affil[1]{Stanford University/Stanford Health Care, Radiology Department, Stanford, CA 94305, USA}
\affil[2]{University of Münster, Computer Vision and Machine Learning Systems, Heisenbergstraße 2, Münster 48149, Germany}
\affil[3]{Google, AI Infrastructure, Mountain View, CA 94043, USA}
\affil[4]{University of Illinois Urbana-Champaign, School of Information Sciences, Champaign, IL 61820, USA}

\affil[*]{Corresponding authors: Joy T. Wu  (joytywu@stanford.edu) and Daniel Beckmann (d\_beck26@uni-muenster.de)}
\affil[$\dag$]{These first co-authors contributed equally to this work}
\affil[$\dag$$\dag$]{These co-senior-authors contributed equally to this work}

\begin{abstract}
[18F]FDG-PET/CT is a cornerstone imaging modality for guiding oncology therapies, yet human expert shortages necessitate more efficient diagnostic aids. While standalone AI models for automatic lesion detection exist, clinical translation remains hindered by AI explainability, reliability, and workflow integration. Meanwhile, human-computer-interaction in radiology remain limited to keyboard, mouse and voice, ignoring experts' faster, natural gaze signal. We present GazeXPErT, a 4D eye-tracking dataset with annotated expert decision windows for tumor detection and measurement on 346 dual-read FDG-PET/CTs. The dataset contributes 9,030 gaze-to-lesion trajectories derived from 3,948 minutes of 60 Hz eye-tracking data, rendered in COCO-style format. GazeXPErT captures experts' visual reasoning patterns when adjudicating suspicious lesions. It aims to facilitate development of trusted, explainable and interactive AI models through understanding expert gaze patterns. Baseline feasibility experiments suggest salient signal is extractable from routinely collected expert gaze (3D nnU-Net Dice: 0.6819 versus 0.6008 without), that gaze-trained vision transformers may aid dynamic lesion localization (74.95\% predicted gaze closer to tumor), and that experts' intent may be predictable from raw gaze (Accuracy 67.53\%, AUROC 0.747).
\end{abstract}

\begin{document}

\flushbottom
\maketitle

\thispagestyle{empty}


\section*{Background \& Summary}

18F-fluorodeoxyglucose Positron Emission Tomography/Computed Tomography (FDG-PET/CT) offers sensitive identification of metabolically active tumors and has become a clinically crucial volumetric imaging modality for cancer staging and treatment response monitoring\textsuperscript{\cite{parihar2023fdg, wahl2009recist, jadvar2017appropriate, czernin2002clinical, weber2005use, lardinois2003staging,boellaard2015fdg, avril2005monitoring, fletcher2008recommendations,delgado2003meta,gregoire2010pet,thorwarth2012integration}}. The PET imaging component uses a radioactively labeled glucose analog to image cells with higher metabolic activities, while CT images provide the corresponding anatomical details to increase diagnostic specificity\textsuperscript{\cite{hofman2016we}}. Reproducible measurement of index and total tumor burden is important for prognosis\textsuperscript{\cite{zhang2013prognostic,meignan2021total}}, but is time consuming to do manually, a burden compounded by a persistent radiologist workforce shortage relative to growing imaging demand\textsuperscript{\cite{christensen2025projected}}. To accurately interpret tumor burden, human experts employ extensive prior knowledge, including anatomy-specific search patterns and cancer-specific metastatic pathways, to differentiate tumors from normal physiological FDG uptake, post-treatment changes, or alternative diagnoses. These skills are taught interactively between experts of differing experience and individually refined over years of training, allowing confident interpretation even in cases very different from prior experience. Crucially, human experts also learn to appropriately express uncertainty when the available information does not make sense. 

In contrast, current state-of-the-art deep learning models trained to segment or detect tumors in FDG-PET/CT scans are static, initially developed on fixed and often small datasets, and trained without experts' visual reasoning insight\textsuperscript{\cite{gatidis2024results,grossiord2017automated,eyuboglu2021multi,gatidis2022whole}}. Although there are techniques to train more robust models through better representation of intrinsic imaging features using self-supervision, attention mechanisms, ensemble models, or contextual multi-modal and multi-tasking models\textsuperscript{\cite{xia2025comprehensive,salimi2024explainable}}, they remain standalone AI models that vary in performance and struggle with real life out-of-distribution data. These barriers, along with concern for model bias, interpretability, and explainability, render most research models inadequate for clinical adoption\textsuperscript{\cite{mbakwe2023fairness,sundar2024automatic,recht2020integrating}}. Moreover, efficient solutions to integrate automatic AI output into the workflow of imaging specialists, without introducing additional cognitive burden, is still very much an open problem\textsuperscript{\cite{droguet2024automated,frood2024comparative,khosravan2019collaborative}}. Notably, most medical imaging AI is still built for an archaic keyboard-mouse-voice interaction paradigm; gaze offers a faster, more natural, dynamic input that modern AI architectures could leverage to improve adoption via more efficient human-AI collaboration.

Our dataset, GazeXPErT, aims to help address these problems by simultaneously capturing radiology experts' visual reasoning patterns and their task intentions. Human eye movements are task specific and are controlled by the fastest muscles in the body\textsuperscript{\cite{hayhoe2014modeling,foulsham2015eye}}. Despite being radiologists' primary sensory modality, gaze remains underutilized in radiology human-AI interactions. Historically, the majority of eye-tracking research in radiology has focused on exploratory analyses of gaze pattern differences, e.g. by experience or fatigue\textsuperscript{\cite{van2017visual,waite2019analysis}}. However, there have been growing interests to integrate eye-tracking with machine learning to develop more trusted and explainable AI models in both the medical and non-medical domains, for example by mimicking human visual attention\textsuperscript{\cite{mall2018modeling,karargyris2021creation,saab2021observational,wang2025improving,zhou2021using,cho2024integration}} and broadening our understanding of the cognitive processes for complex tasks\textsuperscript{\cite{zhang2022can,hessels2019eye,rizzo2022machine}}. Overall, there is a growing body of research in radiology utilizing eye-tracking data to train more interpretable models\textsuperscript{\cite{duan2025eye,moradizeyveh2024eye}}. Notable examples include: training chest X-rays diagnostic or reporting models with radiologists’ search patterns \textsuperscript{\cite{bigolin2022reflacx,pham2024fg,ma2024eye,wang2024gazegnn,zhu2022gaze,hsieh2024eyexnet}}, gaze assisted lesion or anatomy segmentation for CT brain and body\textsuperscript{\cite{khaertdinova2024gaze,stember2019eye,khosravan2016gaze2segment,kirillov2023segment,ma2024segment}}, and radiologists scan path prediction for chest X-rays and CT abdomen\textsuperscript{\cite{pham2025ct}}. 

GazeXPErT is a multimodal 4D eye-tracking dataset that captures 13 independent human experts' search patterns while detecting tumors in 346 whole-body FDG-PET/CTs from a public dataset\textsuperscript{\cite{gatidis2022whole}}. GazeXPErT contributes 9,030 unique gaze-to-lesion trajectories derived from 3,948 minutes of raw eye-tracking data (60 Hz), rendered in a COCO-style format for machine learning tasks\textsuperscript{\cite{rostianingsih2020coco}}. All imaging visualization parameters were recorded simultaneously, enabling reconstruction of the entire FDG-PET/CT read. GazeXPErT is designed to explore multiple machine learning problems, including visual grounding or causal reasoning, clinically explainable feature augmentation, human-computer interaction, human intention prediction/understanding, and expert gaze-rewarded modeling. The dataset will be open-sourced via multiple venues, including the Stanford Center for Artificial Intelligence in Medicine and Imaging (AIMI Center) and Kaggle. 

This paper details our dataset collection protocol and data schema. We show baseline validation experiments, exploring potential benefits of training AI models with expert insight and the technical feasibility of training models to predict task-specific expert intentions through their gaze for FDG-PET/CT. Our goal is to motivate clinical and machine learning communities to reimagine diagnostic medical imaging workflows, where human experts and AI can more adaptively leverage each other's insight in real time.

\section*{Methods}



In clinical FDG-PET/CT interpretation, lesion detection is inherently context and decision-driven rather than purely exhaustive, consistent with classic findings that diagnostic errors in radiology arise from a mix of search, recognition, and decision-making failures\textsuperscript{\cite{kundel1978visual}}. Expert readers prioritize FDG-avid lesions differently depending on the clinical context, such as baseline staging versus treatment response assessment, identification of index lesions, and estimation of total tumor burden, and this higher-order clinical reasoning evolves dynamically during interpretation\textsuperscript{\cite{krupinski2010current}}. In routine interpretations, experts often implicitly prioritize so-called “index lesions” that are most representative of disease activity or most likely to influence management, while de-emphasizing small or clinically inconsequential findings, a prioritization reflected in gaze dwell time and revisit frequency. GazeXPErT's data collection is designed to capture each of these three stages: search, as described above, and recognition and decision-making, addressed in turn next.

Gaze location alone does not guarantee recognition: even expert observers can look directly at a target and fail to perceive it\textsuperscript{\cite{drew2013invisible}}. GazeXPErT therefore does not rely on gaze dwell or fixation as a proxy for recognition; a lesion is only recorded once a reader takes a deliberate, timestamped “select” action followed by an explicit accept, reject, or resize decision, marking a window of unequivocal recognition rather than passive visual search (see Dataset Limitations).

Recognition, in turn, is not the same as deciding why a hot spot is malignant: distinguishing true tumor from physiologic or benign uptake (e.g., brown fat, reactive lymph nodes, inflammatory change) requires cross-referencing PET intensity against CT anatomy, not fixation location alone\textsuperscript{\cite{bruno2015understanding}}. GazeXPErT captures this decision-making process by recording each reader's modality switching, windowing, and scrolling behavior in sync with gaze (see Data Collection), reconstructing not only where a reader looked but how they interrogated the imaging to reach a diagnostic decision. A key open research question this dataset is designed to support is learning which combinations of these recorded viewing behaviors are predictive of a true-tumor determination — moving beyond spatial gaze localization toward modeling the interpretive reasoning process itself.

\subsection*{Ethics Declarations}

IRB approval (\#82421) was obtained from Stanford Health Care/Stanford School of Medicine to add expert eye-tracking insight and new independent lesion annotations to an existing de-identified public dataset\textsuperscript{\cite{gatidis2022whole}} for interpretability, explainability, reasoning, and interactive AI modeling. The research project also obtained a data use agreement from Tübingen for usage of the FDG-PET/CT images for eye-tracking AI interpretability and explainability research.

\subsection*{Data Collection}
\begin{figure}[ht!]
    \centering
    \includegraphics[width=0.5\linewidth]{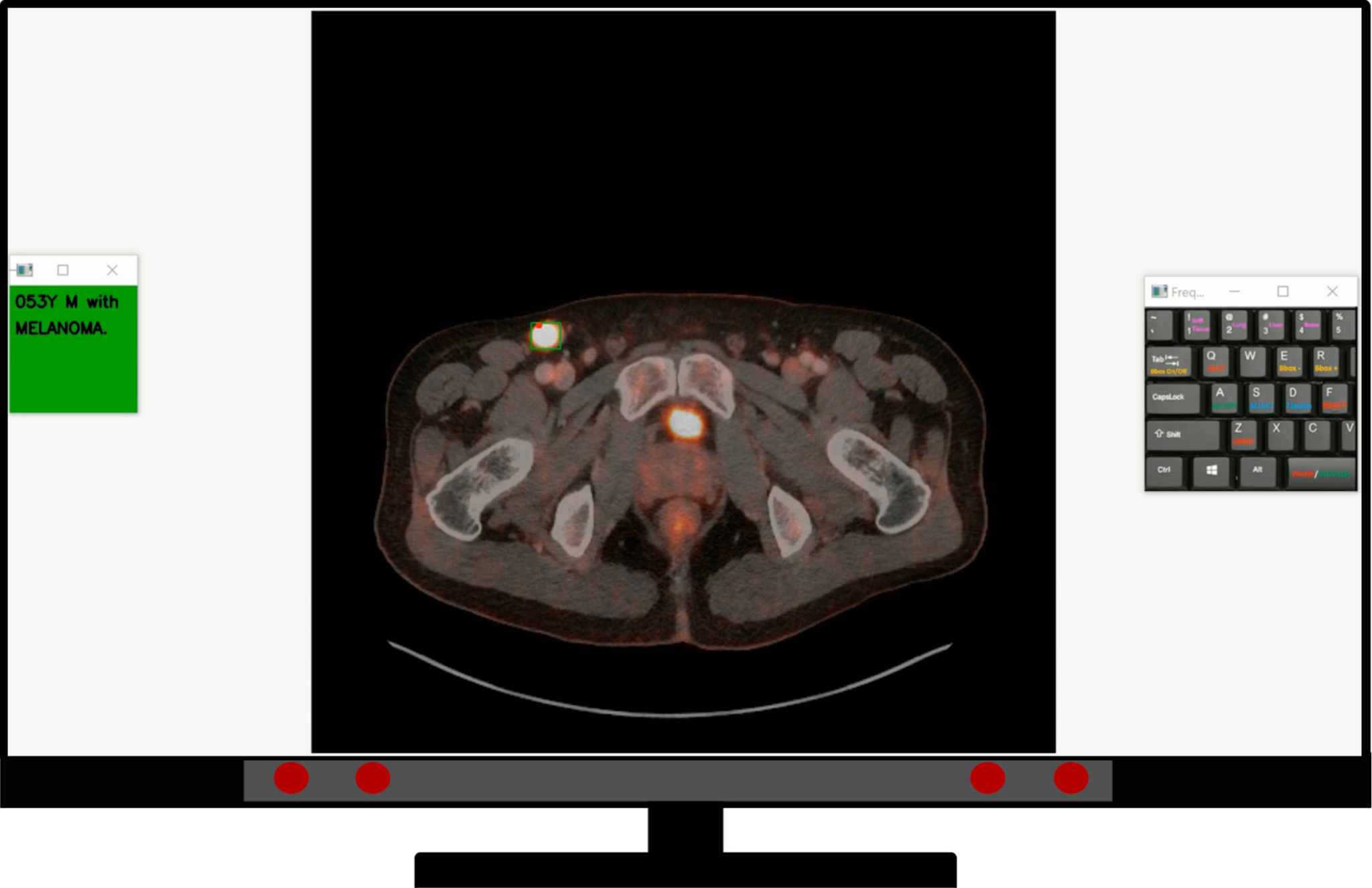}
    \caption{Data collection platform. The PET/CT study was shown to readers in a full-screen application alongside some information about the patient and a keybindings legend.}
    \label{fig:annotation_tool}
\end{figure}

A custom eye-tracking PET/CT platform (Figure~\ref{fig:annotation_tool}) was built with Python packages to record the search patterns and lesion annotations of human experts while reading FDG-PET/CT studies. 346 unique studies (98 lymphoma, 122 lung cancer, and 126 melanoma) from 343 patients were randomly sampled from an open-source whole-body FDG-PET/CT dataset\textsuperscript{\cite{gatidis2022whole}}. Each study was re-interpreted by one of six trainees (five radiology residents and one nuclear medicine fellow) and one of seven experienced board-certified specialists in nuclear medicine and/or diagnostic radiology (13 unique readers in total). Trainees were all based at Stanford University and had 2-3 years of clinical experience at the time of data collection. Experienced readers were based at Stanford University and partner institutions in Germany, Japan, and Korea, with 2-12 years of post-training clinical experience. Given the patient's age, gender, and cancer diagnosis, the observer's task was to identify, “gaze-select” and confirm all metabolically active lesions, as well as indicate how likely each lesion was related to cancer (“certain” versus “uncertain”). After a one-on-one eye-tracking platform coaching session, every observer reads the same “inter\_observer” practice study at least once before beginning formal data collection. In addition, before each reading session, the readers re-calibrate their gaze to the eye-tracking station’s set up using the manufacturer’s desktop calibration software.

Readers were instructed to simulate a comprehensive clinical interpretation consistent with routine oncologic PET/CT practice, prioritizing identification of metabolically active lesions relevant to disease staging and overall tumor burden assessment rather than a single predefined clinical endpoint. The inclusion of a “certain” versus “uncertain” lesion label reflects routine clinical PET interpretation, where many FDG-avid findings—particularly small lymph nodes or mildly hypermetabolic foci—cannot be definitively classified without follow-up imaging or additional clinical information. Rather than representing annotation noise, this uncertainty encodes expert diagnostic judgment and confidence, which is a critical but often unmodeled component of real-world oncologic image interpretation. Capturing this uncertainty enables future machine learning models to learn not only where experts look, but how confidently they interpret findings.  


Simulated routine PET/CT read set up: A screen-based research eye-tracking device (Tobii Pro Spark\textsuperscript{\cite{TobiiProSpark}}) was used to track a reader's eye movement and pupil sizes, recording corresponding 2D gaze locations on a diagnostic 27-inch screen at a sampling frequency of 60 Hz. Participants were allowed to adjust image contrast and windowing, scroll back and forth between 2D axial image slices (i.e. reasoning through the 3D volume), and switch between three different “modalities” (PET, CT, fused PET/CT) to identify and confirm malignant lesions. For the PET modality, readers had the additional option to view the volumetric data in a rotatable upright whole-body Maximum Intensity Projection (MIP) from 12 angles. All image visualization parameters were recorded at 60 Hz simultaneously with the corresponding gaze coordinates to allow video-like reconstruction of the entire read process. The original 512$\times$512 CT images and the upsampled (to 512$\times$512) and standardized PET images (i.e., SUV) were read for the study. More detailed annotation protocol can be found on our GitLab repository.

\subsection*{Gaze-Assisted Lesion Annotation}

Beyond gaze coordinates, ground truth annotations for tumors were collected at the bounding box level at each image slice. The annotation was accelerated semi-automatically via a candidate proposal and validation process (see Candidate Lesion Proposal Algorithm below), where readers utilized hotkeys to indicate when they wanted to select, accept, reject, or resize a proposed tumor candidate. The proposed bounding box for a lesion could be resized by changing the PET intensity threshold used to generate the candidate via two hotkeys. This PET threshold was recorded and can be used to derive a pseudo-pixel segmentation mask that best estimates the lesion boundary. Notably, readers were instructed to do their best to visually match the PET lesion boundary with the CT lesion boundary when possible. To achieve true recall for the dataset, a baseline candidate lesion proposal algorithm (see details below) was chosen for its high sensitivity so that readers could select almost any suspicious hypermetabolic lesion and not be limited by the misses from an existing AI segmentation model.

\subsection*{Candidate Lesion Proposal Algorithm}

The lesion candidate proposal algorithm operates as follows: the last gaze location upon the “select” keystroke is used as the reference point. A user-adjustable maximal SUV (normalized PET) threshold of a slice is used to create pseudo-masks via maximal intensity thresholding, and boxes around masks are calculated via 2D connected components analysis with 8 connectivity. Previously rejected and accepted boxes are filtered out, and the box nearest to the last gaze location is proposed and overlaid over the image slice for display. The size of the box can be adjusted in real-time by increasing or decreasing the SUV segmentation threshold, and the box gets accepted or rejected (for example, because a wrong box is proposed due to inaccurate last gaze point). All key presses are timestamped and saved to discretely record a reader's interactive decisions throughout a study read.

\subsection*{Bounding Box Propagation and Annotation Efficiency}

To save time and effort, an accepted box on a “root” (current) slice gets propagated into neighboring slices. Neighboring slices in both z-directions are processed using the saved SUV threshold from the root “accepted” slice. The reader can additionally correct the sizes of the boxes on the adjacent slices separately. The efficiency of this approach is reflected in the annotation time metrics: over 49 hours of gaze data were collected (following preprocessing of the full 3,948 minutes, 65.8 hours, of raw eye-tracking capture to remove periods when the reader was not looking at the screen), with a median annotation time per study of 191 seconds and a mean of 259 seconds. The median annotation time per 3D lesion was 0.81 seconds, with a mean of 1.51 seconds. This demonstrates the feasibility of our gaze-assisted annotation approach for similar large-scale dataset collection. Based on key press logs, approximately 19\% of accepted lesion bounding boxes and 15\% of rejected lesion candidates were subsequently resized or otherwise modified by the reader, indicating that most candidate proposals required little manual correction; these ratios are keystroke-derived estimates and may carry some imprecision.


\section*{Data Records}
The GazeXPErT dataset contains no identifying information and will be hosted on multiple platforms, including the Stanford Center for Artificial Intelligence in Medicine and Imaging (AIMI Center) and Kaggle. While the paper is undergoing peer review, the dataset will only be hosted by the Stanford AIMI Center at: \url{https://stanford.redivis.com/datasets/460a-2j4r7trxr}.


\subsection*{Dataset Structure Overview}
\begin{figure}[ht!]
\centering
\begin{subfigure}[b]{.33\textwidth}
  \centering
  \includegraphics[width=1\linewidth]{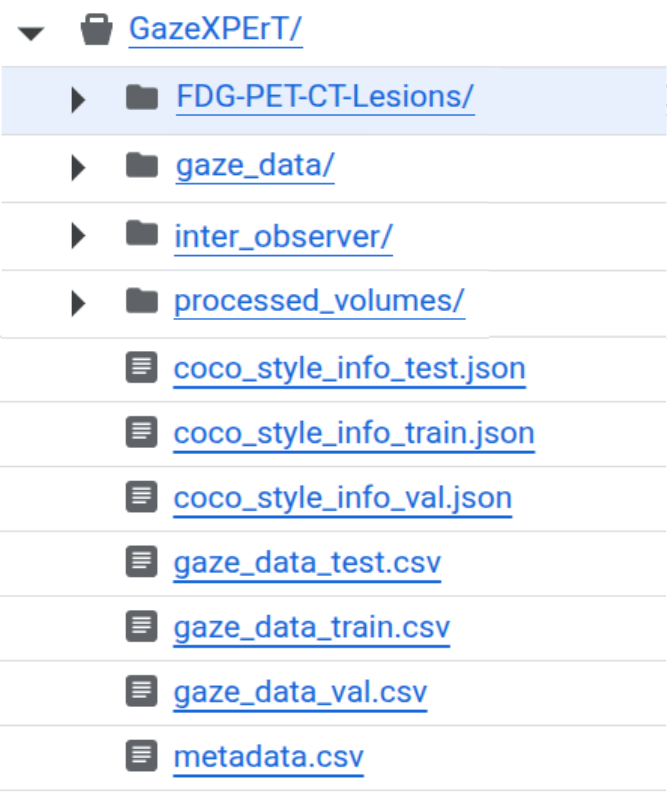}
  \caption{Dataset folder structure.}
  \label{fig:folder_structure}
\end{subfigure}%
\begin{subfigure}[b]{.66\textwidth}
  \centering
  \includegraphics[width=\textwidth]{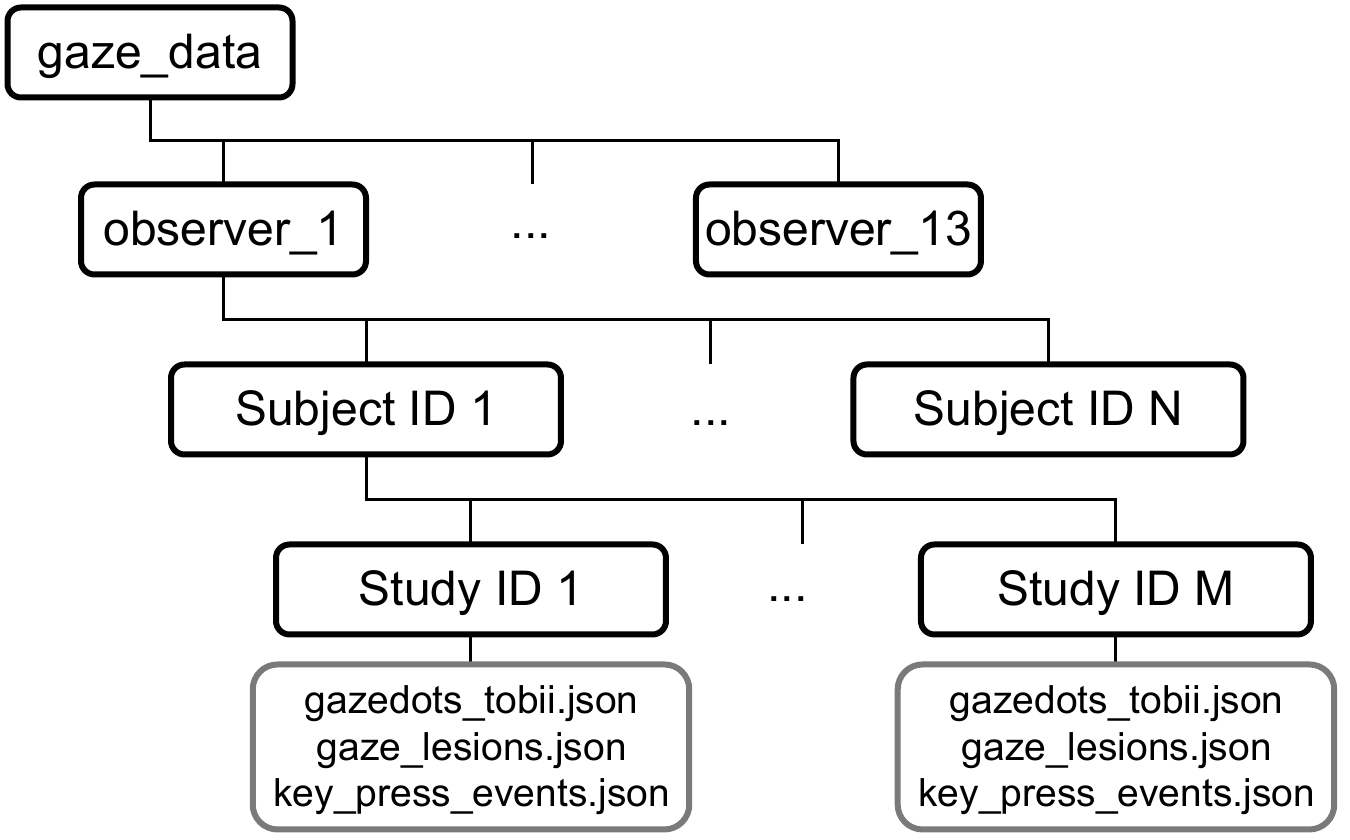}
  \caption{Structure of raw gaze recordings.}
  \label{fig:gaze_data_structure}
\end{subfigure}%
\caption{GazeXPErT dataset structure overview. Figure~\ref{fig:folder_structure} depicts the top-level folder structure, Figure~\ref{fig:gaze_data_structure} illustrates the structure of raw recordings, following the Tübingen dataset structure.} 
\label{fig:structure_overview}
\end{figure}

The GazeXPErT dataset consists of two sets of added expert visual insight (gaze trajectories) and lesion-level annotations in the form of raw JSONs and postprocessed gaze segmentation volumes indexed to the source dataset. The original PET/CT images came from the open-source Tübingen “autoPET” whole-body FDG-PET/CT dataset\textsuperscript{\cite{gatidis2022whole}}, and both datasets are shared under the Creative Commons Attribution 4.0 International License (CC BY 4.0). The same source imaging data also underlies the autoPET Challenge series\textsuperscript{\cite{gatidis2024results,dexl2026autopet}}; we used the original public release of the Tübingen FDG-PET/CT collection as described in Gatidis et al.\textsuperscript{\cite{gatidis2022whole}}, with no subsequent re-versioning applied on our end. The highest level structure of the GazeXPErT repository is organized as in Figure~\ref{fig:structure_overview}, where the highlighted “FDG-PET-CT-Lesions” folder will need to be separately downloaded from Tübingen (see setup instructions from our GitLab repository).

The “gaze\_data” folder contains the raw eye-tracking data in the GazeXPErT dataset. The “inter\_observer” folder contains at least 1 practice recording for each observer, which is also used for deriving gaze calibration measurement data. The “processed\_volumes” folder contains the eye-tracking data rendered in 3D heatmap volume, useful for 3D modeling. The root “gaze\_petct” folder also contains the “metadata.csv”, which retains the information to link recorded gaze and annotations to their respective original anonymized studies by study paths.

As shown in Figure~\ref{fig:gaze_data_structure}, the structures for both the “gaze\_data” and the “inter\_observer” folders are first organized by anonymized observer IDs, then by the original nested study paths. Each study recording folder contains three separate files for raw gaze data (“gazedots\_tobii.json”), lesion annotations (“gaze\_lesions.json”) and key presses (“key\_press\_events.json”) respectively. The observer ID level is removed from the structure of the “processed\_volumes” folder.


\subsection*{Raw Gaze Data JSONs Explained}
For each study read, there are three types of raw JSON files, detailed below. Code to process and utilize all the gaze JSON and CSV information is provided in our GitLab.

\begin{figure}[ht!]
    \centering
    \includegraphics[width=\linewidth]{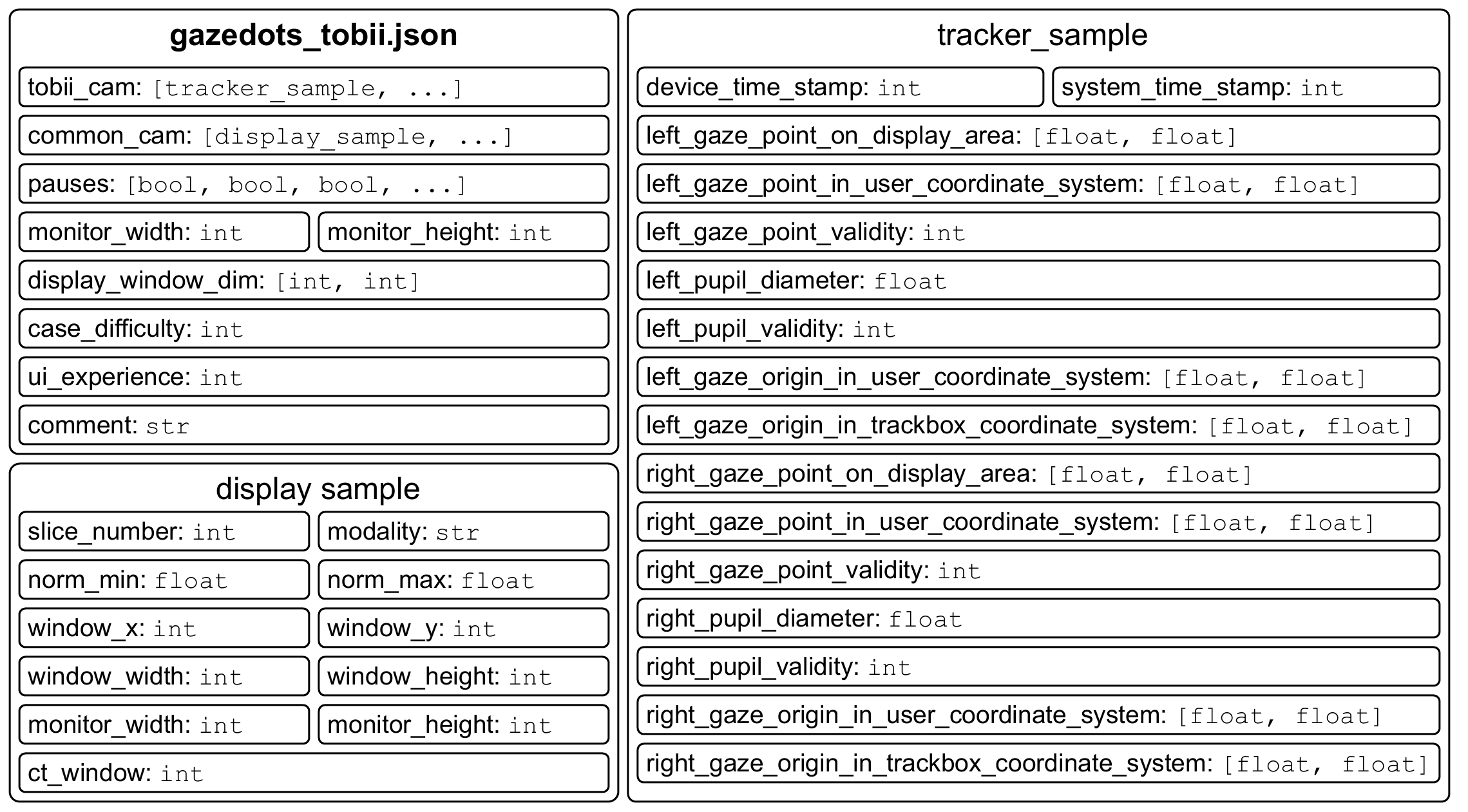}
    \caption{Content overview for “gazedots\_tobii.json”.}
    \label{fig:gaze_json_description}
\end{figure}

\paragraph*{“gazedots\_tobii.json” explanation (Figure~\ref{fig:gaze_json_description}):}
\begin{itemize}
\itemsep0em 
\item Each file contains the raw gaze and image metadata recording for the entire study read from one observer.
\item The fields “tobii\_cam”, “common\_cam” and “pauses” contain elements synced by position in ordered lists.
\item The “tracker\_sample” shows the raw tobii data saved for each gaze point. For more information, please see user support for Tobii Pro Spark tracker.
\item The “display\_sample” shows the corresponding imaging metadata saved with each gaze point (at 60 Hz too) to reconstruct the image in the way that the experts deemed best to visualize the PET/CT data at that point in time.
\item The “pause” tag shows when the reader decided to take a break from reading, which is uncommon besides near the start and end of each read. It is also by default set to “True” in the time between a lesion “select” and its “accept” or “reject” decision key presses to tag the gaze within this time window, as there can be a small computational lag to propagate the lesion candidate bounding box annotations to image slices adjacent to the root image slice.
\end{itemize}

\begin{figure}[ht!]
\centering
\begin{subfigure}[b]{.45\textwidth}
  \centering
  \includegraphics[width=\textwidth]{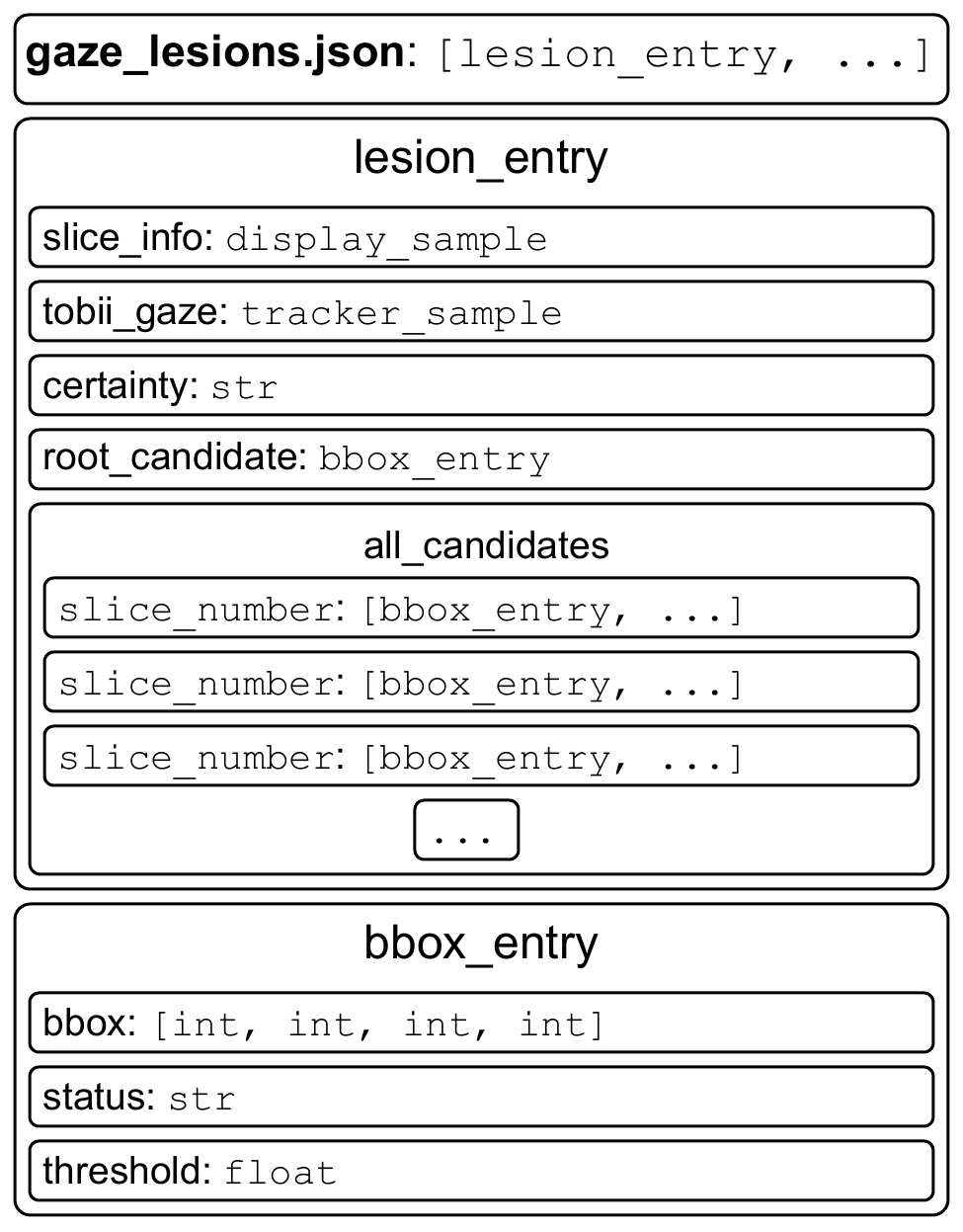}
  \caption{}
  \label{fig:lesion_json}
\end{subfigure}%
\begin{subfigure}[b]{.45\textwidth}
  \centering
  \includegraphics[width=\textwidth]{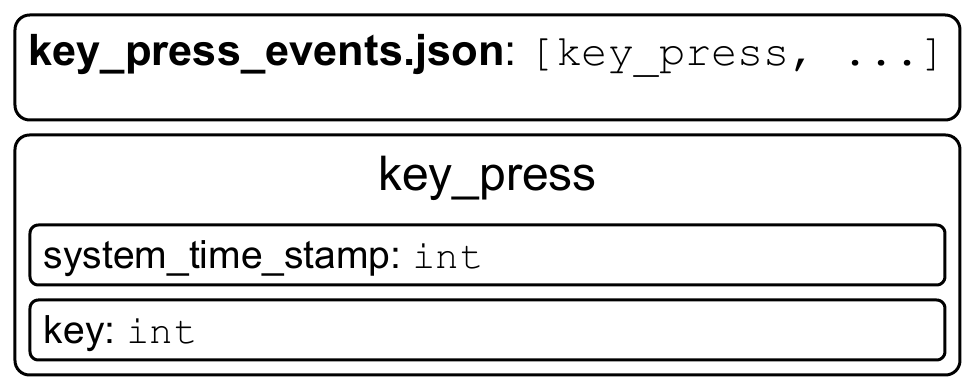}
  \caption{}
  \label{fig:keypress_json}
\end{subfigure}%
\caption{Content overview for “gaze\_lesions.json” (a) and “key\_press\_events.json” (b) files, respectively. These files contain information regarding annotated lesions and all keystrokes during the annotation process.}
\label{fig:other_jsons}
\end{figure}
\FloatBarrier

\paragraph{“gaze\_lesions.json” explanation (Figure~\ref{fig:lesion_json}):}
\begin{itemize}
\itemsep0em 
\item Each file contains the subset of raw gaze recording associated with the “root” lesion annotations at the bounding box (Bbox) level. The Bboxes are stored in the “xywh” format in pixel coordinates using the display size during annotation (e.g., if displayed at 1400px, then the coordinates would be upscaled to 1400 px from the original PET/CT images’ 512 px resolution).
\item Each lesion annotation contains: the raw gaze coordinate at the point of lesion selection, the root accepted Bbox candidate coordinates, the candidate acceptance key used (“certain” versus “uncertain”), the corresponding image screen display parameters, all the Bboxes from all adjacent slices associated with the root lesion slice, and the PET “threshold” used to create them.
\item “certainty”: The observer can use two different acceptance keys to indicate how confident they are that the lesion is due to a tumor (“certain”) versus another possible explanation but cannot exclude cancer (“uncertain” for e.g., metastatic versus reactive lymph node).
\item “status”: Whether the lesion bounding box on a slice was manually validated versus extrapolated candidates from a root-validated/accepted lesion.
\end{itemize}

\paragraph{“key\_press\_events.json” explanation (Figure~\ref{fig:keypress_json}):}
\begin{itemize}
\itemsep0em 
\item This file allows for complete reconstruction of all the manual decisions made during the whole annotation process, including corrections, redactions, etc.
\item The system timestamps sync key press events with the gaze data. All key press events (i.e., experts' decisions) used by the observer for a study read are saved as integers recorded by the OpenCv Python library.
\item Scrolling through the volume using the mouse wheel is not recorded in this file, but can be reconstructed from image slice information in the “display\_samples” field from “gaze\_lesions.json”.
\end{itemize}


\Needspace{6\baselineskip}
\subsection*{Post-processed Gaze Data for Ease of Machine Learning Tasks}
To facilitate ease of using this rich but complex medical imaging eye-tracking dataset, we have postprocessed the gaze data to 1) 3D gaze volume NIFTI files and to a 2) Common Objects in Context (COCO) style format JSONs and CSVs for different types of machine learning tasks.

\paragraph{3D “processed volumes” NIFTI files} for ease of 3D lesion segmentation-related experiments.  These are kept in the “processed\_volumes” folder under their relevant study paths, where * in the filenames below are replaced with either “trainee” or “experienced”.
\begin{itemize}
\itemsep0em 
\item Processing for the 3D Gaze heatmap volume (“GAZE\_*.nii.gz”) and MIP (“GAZE\_MIP\_*.nii.gz”) NIFTI files: Raw gaze trajectory was transformed into 3D volume with the same image shape as the original 512$\times$512 CT study. Each voxel counts the number of gaze samples on it. The temporal component is flattened and no blurring of the aggregate data was used to keep the information “raw”. Maximal Intensity Projection (MIP) of 12 different angles of the 3D gaze heatmap volume is also generated for whole-body static analysis. 
\item Processing for the pseudo segmentation volume NIFTI files (“PSEUDO\_SEG\_*.nii.gz”): 3D segmentation labeled mask derived using the recorded SUV max threshold from the Bbox lesion annotation.
\end{itemize}

\paragraph{COCO-style format JSON and CSV files}, for ease of lesion-level modeling tasks, are kept at the first level immediately below the root “gaze\_petct” folder. 
\begin{itemize}
\itemsep0em 
\item These files organize access to lesion data in the GazeXPErT dataset in a COCO-style format for machine learning experiments. There are 3 JSON (“coco\_style\_info\_*.json”) and 3 CSV (“gaze\_data\_*.csv”) files, organizing annotations along the “train”, validation (“val”), and “test” splits ($\sim$80/10/10) used in our validation experiments. Splits were generated at the patient level (all studies and lesions from a given patient assigned to a single split) to prevent data leakage across train, validation, and test sets.
\item The GazeXPErT dataset consists of gaze-to-lesion trajectories, which can be reconstructed into videos using the source PET/CTs and our synced metadata. These processed gaze data JSON and CSV files capture 9,030 accepted and 2,658 rejected gaze-to-lesion trajectories.
\item In addition, the COCO-style JSONs allow GazeXPErT annotations to be accessible on three levels of detail. Study-read-level links the annotator (observer) and raw gaze data files to the PET/CT image data. Lesion-level (i.e. object-level) aggregates the gaze data to unique lesion levels, and links individual index lesions to relevant temporal gaze data in the separate train, val, or test CSV file. Slice-Level allows each lesion Bbox to be indexed to and retrieved separately.
\item Figure~\ref{fig:coco_json_structure} shows the first (simplified) entry of each field in “coco\_style\_info\_*.json” for one example study read, and Figure~\ref{fig:gaze_csv_example} shows corresponding, correlated rows from “gaze\_data\_*.csv”. The “annotations” list (slice-level) is closest to a standard COCO per-instance annotation object (\texttt{id}, \texttt{image\_id}, \texttt{bbox}); the “lesion\_annotations” list (object-level) is the GazeXPErT equivalent of a COCO “object”, aggregating the root 2D \texttt{bbox}, its 3D extension \texttt{bbox\_3d} across adjacent slices, and links to the corresponding temporal gaze data. Since GazeXPErT annotates a single object class (“lesion”), no \texttt{category\_id} is needed, and bounding boxes are stored in place of pixel-level COCO \texttt{segmentation} masks.

\begin{figure}[ht!]
\centering
\begin{subfigure}[t]{.47\textwidth}
\centering
\begin{lstlisting}[style=jsonstyle]
{
  "info": {
    "version": "0.1",
    "description":
     "Gaze annotation on a single
      FDG-PET/CT scan from an
      anonymized reader.",
    "date_created": "2025-12-11"
  },
  "licenses": [
    {"id": 0, "name": "CC BY 4.0"}
  ],
  "volumes": [{
    "id": "study_001",
    "folder_path":
     "FDG-PET-CT-Lesions/study_001"
  }],
  "images": [{
    "id": "study_001.458",
    "width": 512, "height": 512,
    "file_name":
     "slice_images/study_001.458.jpg"
  }],
  "annotations": [{
    "id":
     "study_001.observer_12.0.458.0",
    "image_id": "study_001.458",
    "bbox": [235, 257, 8, 11],
    "status": "accepted",
    "threshold": 13.51,
    "observer": "observer_12",
    "experience_level": "trainee"
  }]
}
\end{lstlisting}
\caption{}
\label{fig:coco_json_structure_a}
\end{subfigure}%
\hfill
\begin{subfigure}[t]{.47\textwidth}
\centering
\begin{lstlisting}[style=jsonstyle]
{
  "volume_annotations": [{
    "id": "study_001.observer_12",
    "volume_id": "study_001",
    "gaze_filename":
     "gaze_data/observer_12/
      study_001/gazedots_tobii.json",
    "annotation_filename":
     ".../gaze_lesions.json",
    "keypress_filename":
     ".../key_press_events.json",
    "observer": "observer_12",
    "experience_level": "trainee"
  }],
  "lesion_annotations": [{
    "id": "study_001.observer_12.0",
    "volume_id": "study_001",
    "root_annotation": {
      "image_id": "study_001.458",
      "bbox": [235, 257, 8, 11],
      "status": "accepted",
      "threshold": 13.51
    },
    "bbox_3d": [234, 253, 451,
                10, 18, 10],
    "view_parameters": {
      "slice_number": 458,
      "modality": "fused", ...
    },
    "gaze_time_stamp": 1049870,
    "certainty": "certain",
    "observer": "observer_12",
    "experience_level": "trainee"
  }]
}
\end{lstlisting}
\caption{}
\label{fig:coco_json_structure_b}
\end{subfigure}
\caption{First (simplified) entry of each field in “coco\_style\_info\_*.json”, all drawn from the same example study read, observer, and index lesion. \textbf{(a)} File-level metadata (\texttt{info}, \texttt{licenses}) and study/image-level records (\texttt{volumes}, \texttt{images}), plus the slice-level \texttt{annotations} entry for the lesion's root (index) slice. \textbf{(b)} The study-read-level \texttt{volume\_annotations} entry (linking the observer's raw gaze/annotation/keypress files to the volume) and the object-level \texttt{lesion\_annotations} entry for the same lesion. Identifiers are simplified/generalized from the real anonymized values for readability (e.g. “study\_001” in place of the full source study path); numeric fields (bbox, threshold, slice number, timestamp) are real values from the “test” split. “...” denotes fields omitted for space (e.g. remaining \texttt{view\_parameters}, repeated folder paths).}
\label{fig:coco_json_structure}
\end{figure}

\item Figure~\ref{fig:gaze_csv_example} correlates a few rows of “gaze\_data\_test.csv” with the \texttt{lesion\_annotations} entry from Figure~\ref{fig:coco_json_structure}(b), showing how the temporal gaze stream links to a specific lesion annotation event.

\begin{figure}[ht!]
\centering
\begin{tabular}{
>{\raggedleft\arraybackslash}p{2.3cm}|
>{\centering\arraybackslash}p{1.9cm}|
>{\centering\arraybackslash}p{3.6cm}|
>{\centering\arraybackslash}p{3.6cm}|
>{\centering\arraybackslash}p{2.4cm}|
>{\centering\arraybackslash}p{1.2cm}}
    \hline
    \texttt{\scriptsize device\_time\_\newline stamp} & \texttt{\scriptsize slice\_number} & \texttt{\scriptsize center\_gaze\_point\_\newline on\_pixel\_image\_x} & \texttt{\scriptsize center\_gaze\_point\_\newline on\_pixel\_image\_y} & \texttt{\scriptsize left\_pupil\_\newline diameter} & \texttt{\scriptsize paused} \\
    \hline
    \hline
    1000000 & 458 & 239 & 272 & 3.60 & False \\
    1016613 & 458 & 240 & 272 & 3.59 & False \\
    1033653 & 458 & 240 & 271 & 3.58 & False \\
    \textbf{1049870} & 458 & 240 & 272 & 3.58 & \textbf{True} \\
    1066527 & 458 & 239 & 275 & 3.57 & True \\
    1083573 & 458 & 240 & 276 & 3.56 & True \\
    1099836 & 458 & 242 & 276 & 3.55 & True \\
    \hline
\end{tabular}
\caption{Example consecutive rows from “gaze\_data\_test.csv” (real values, test split), all sharing \texttt{volume\_annotation\_id} = “study\_001.observer\_12” and \texttt{slice\_number}~=~458, i.e. the same study read and root slice as Figure~\ref{fig:coco_json_structure}. \texttt{device\_time\_stamp} values are shown as simplified relative device-clock values (shifted to a round base for readability) but preserve the true $\sim$16.7\,ms (60 Hz) sampling interval between rows. The bolded row's timestamp (1049870) exactly matches \texttt{gaze\_time\_stamp} in the Figure~\ref{fig:coco_json_structure}(b) \texttt{lesion\_annotations} entry: this is the instant the reader pressed the lesion “select” key, after which \texttt{paused} flips to “True” (consistent with the pause-tagging behavior described in Raw Gaze Data JSONs Explained). At that instant, the gaze pixel coordinates (240, 272) fall close to the annotated bbox center ($\approx$239, 262), consistent with the gaze accuracy reported in Technical Validation.}
\label{fig:gaze_csv_example}
\end{figure}

\item Gaze postprocessing for the CSV files: The “gaze\_data\_*.csv” files contain the extra temporal eye-tracking data dimension that does not fit within the standard COCO style JSON format. Again, the gaze data across studies are aggregated into one csv per train, val, test split, same as for the COCO-style format JSONs. Raw gaze from Tobii tracker is filtered to exclude invalid samples (viewing outside the annotation window or flagged invalid by Tobii). Each row in the CSV maintains the link to the source FDG-PET/CT study read (via “volume\_annotation\_id”) and the slice image, and stores the timestamps, the corresponding gaze coordinates (raw 3D Tobii gaze point/origin, and 2D coordinates rescaled to the original 512px axial PET/CT images), and left/right pupil diameters. Lesion Bbox coordinates are not duplicated in the CSV; they can instead be cross-referenced with the corresponding “annotations” and “lesion\_annotations” entries in “coco\_style\_info\_*.json” by joining on “volume\_annotation\_id” and slice number, as shown in Figure~\ref{fig:gaze_csv_example}.
\end{itemize}
\FloatBarrier

\section*{Technical Validation}

First, to technically evaluate the eye-tracking data quality, we provide both gaze calibration error and inter-reader agreement analyses, and compare our results with the respective existing literature. Second, for use case validation, we show three modeling experiments chosen to represent distinct, broad categories of use for GazeXPErT's gaze and interaction data: 1) 3D FDG-PET/CT tumor segmentation with and without expert insight, testing gaze as an auxiliary signal for a standard downstream AI task; 2) using temporal gaze and image data to improve lesion targeting for the dynamic interactive radiology read setting, testing whether the raw gaze signal itself can be refined for more practical data capture; and 3) demonstrating the feasibility of task-specific expert intentions prediction through their gaze, testing whether reader intent can be inferred to support real-time interactive assistance. These three experiments are representative examples, not an exhaustive account of GazeXPErT's potential applications. We emphasize that they are intended as feasibility demonstrations of the kinds of tasks the GazeXPErT dataset can support, using established, off-the-shelf model architectures without significant task-specific optimization. They are not intended to demonstrate model novelty, cross-institutional generalizability, or readiness for clinical deployment, and should not be interpreted as evidence that gaze-informed AI models currently improve real-world PET/CT workflows.

\subsection*{Gaze Calibration}
As described in the methods section, each observer calibrated the eyetracker before reading using Tobii's supplied proprietary software, which does not report calibration metrics. In addition, since the annotation task involves the observers “gaze-selecting” moving targets (lesions) while they scroll through the PET/CT image slices, we measure the task-specific gaze calibration quality using the same inter-observer study read by all observers. If an observer read the inter-observer study multiple times, the last practice run was used for this analysis.  In contrast to the normal calibration procedure, which utilizes fixed point-like targets, we calculated the calibration metrics on real-world targets, i.e. on annotated lesions in the inter-observer study. Specifically, upon selecting a lesion in the annotation software, we use the gaze samples of the last $0.25$ seconds leading up to the lesion selection to calculate gaze accuracy and precision. We define the center of the resultant lesion bounding box on the current (root) slice as the target point.

Gaze accuracy is defined as the mean deviation of the gaze samples from the target point, and precision is calculated as the root mean square (RMS) between consecutive gaze samples. We report both metrics in terms of degrees angle between the two gaze vectors from the eye position (i.e., the gaze origin) to the recorded gaze sample and the target point, respectively. Across all observers (total N=149 with average 11.4 lesion calibration points per observer), we report a mean gaze accuracy of $1.176^\circ$ (standard deviation of $0.692$), and a mean precision of $0.314^\circ$ (standard deviation of $0.167$) in the last 0.25 seconds.

In comparison to static calibration results, our setup achieved mean precision within the expected range of $0.1^\circ$ to $0.5^\circ$, while mean accuracy lies outside the usually reported range of $0.3^\circ$ to $0.75^\circ$~\textsuperscript{\cite{hooge2024eye}}. The worse mean accuracy may in part be attributed to the nature of our task, which involves selecting dynamic moving targets, as lesions pop in and out of field of view on axial images. i.e. observers are not expected to look directly at the center of the lesion when using our eye-tracking enabled PET/CT annotation software.

Table~\ref{tab:calibration} separately reports the angular distance to the target from two single reference points per lesion — the last gaze sample and the closest gaze sample within this last 0.25 seconds gaze window.

\begin{table}[h!]
    \centering
    \begin{tabular}{r|c|c|c|c}
    \hline
     &Mean (std) & Median & Min & Max \\
     \hline
     \hline
     Last Gaze Point  & 1.068 $\pm 0.672$ & 0.754 & 0.483 & 3.056 \\
     Closest Gaze Point     & 0.837 $\pm 0.648$ & 0.604 & 0.269 & 2.811\\
     \hline
\end{tabular}
    \caption{Distance of the last and closest gaze point to the lesion center, across all participants and lesions.}
    \label{tab:calibration}
\end{table}

\subsection*{Inter-Annotator Agreement}
In order to assess the annotation quality of the GazeXPErT dataset, we compare our two independent overreads to each other and to the original “Tübingen” annotation provided by Gatidis et al.~\textsuperscript{\cite{gatidis2022whole}}. The analyses do not intend to imply which set of ground truth is “more correct” since the annotation tasks and goals between Tübingen and the GazeXPErT dataset are inherently different, and the two datasets are intended to be used synergistically. We analyzed the agreement between the three different sets of lesion annotations at the 3D lesions level. Since our annotation method yields bounding boxes, we transform the segmentation ground truth from the Tübingen dataset accordingly. We calculated the standard average fixed raters Intraclass Correlation Coefficient\footnote{ICC implementation from the Python package \texttt{pingouin}} (ICC) metric, used in the FDG-PET/CT segmentation literature\textsuperscript{\cite{jacene2009assessment,constantino2023evaluation}}. However, since ICC only accounts for matching lesions, we also calculate additional precision, recall and percentage inter-observer agreement ($\frac{\mathrm{TP}}{\mathrm{TP} +\mathrm{FP} + \mathrm{FN}}$) metrics between the three sets of annotations. For the latter analysis, due to predominantly small lesions (objects) in the dataset, a match (TP) is calculated at Intersection over Union (IoU) greater than $0$. Table~\ref{tab:annotation_agreement} summarizes our findings.

Overall, the inter-reader agreement by ICC metrics across all 3 sets of annotations fall within the reported ICC range (0.4-0.9) in the existing literature for FDG-PET/CT lesion measurement\textsuperscript{\cite{jacene2009assessment,constantino2023evaluation}}. Of the matching lesions, GazeXPErT trainees and experienced annotations do agree more with each other than with the Tübingen annotations by Gatidis et al.~\textsuperscript{\cite{gatidis2022whole}}, which is expected given their shared task instruction, setup and overall practice environment. Across all lesion annotations, the GazeXPErT experienced readers do agree more with the Tübingen annotations than trainees, although the difference is not statistically significant at the 95\% confidence interval. The trainees also have more overcalled lesions relative to both the GazeXPErT and Tübingen annotations, predominantly due to small uncertain lesions — this lowers \%Agreement (which penalizes unmatched lesions) without necessarily lowering ICC (computed only on the lesions both readers matched), explaining why Trainee-Experienced has the highest ICC but lowest \%Agreement of the three pairs.


\begin{table}[h!]
    \centering
    \begin{tabular}{rcl|c|c|c|c}
    \hline
     &&& Precision & Recall &  \% Agreement & ICC [95\% CI] \\
     \hline
     \hline
     Trainee &$\leftrightarrow$& Experienced  & 0.8130 & 0.8165 & 0.6748 & 0.76 [0.75-0.78] \\
     Trainee &$\leftrightarrow$& Tübingen     & 0.8355 & 0.8371 & 0.7140 & 0.70 [0.68-0.73] \\
     Experienced &$\leftrightarrow$& Tübingen & 0.8560 & 0.8668 & 0.7600 & 0.74 [0.72-0.76] \\
     GazeXPErT Union &$\leftrightarrow$& Tübingen       & 0.8458 & 0.8519 & 0.7369 & 0.73 [0.71-0.75]\\
     \hline
\end{tabular}
    \caption{Three Way Inter-reader Lesion Annotation Agreement Analysis between GazeXPErT and Tübingen annotations.}
    \label{tab:annotation_agreement}
\end{table}

\begin{figure}[ht!]
\centering
\begin{subfigure}{.33\textwidth}
  \centering
  \includegraphics[width=1\linewidth]{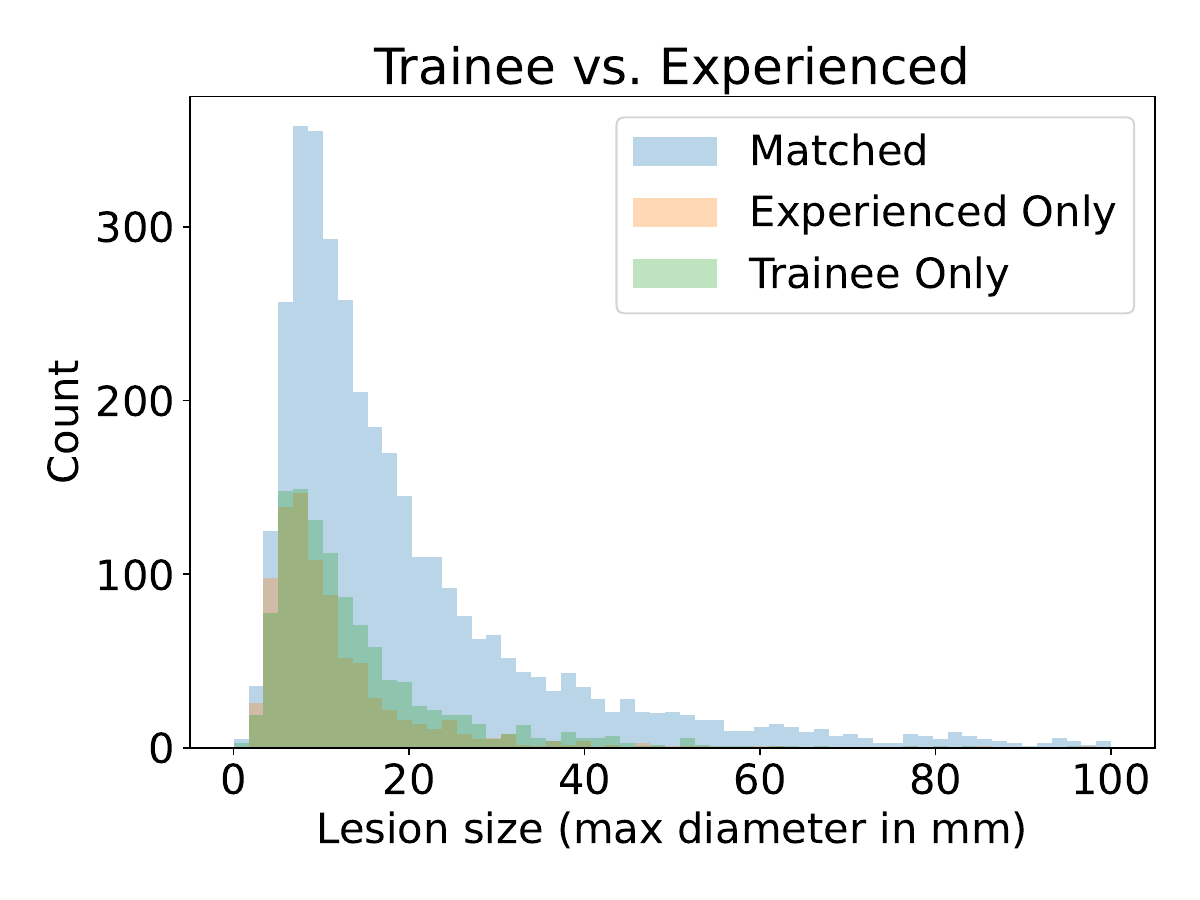}
  \caption{}
  \label{fig:sizecomp_inter}
\end{subfigure}%
\begin{subfigure}{.33\textwidth}
  \centering
  \includegraphics[width=1\linewidth]{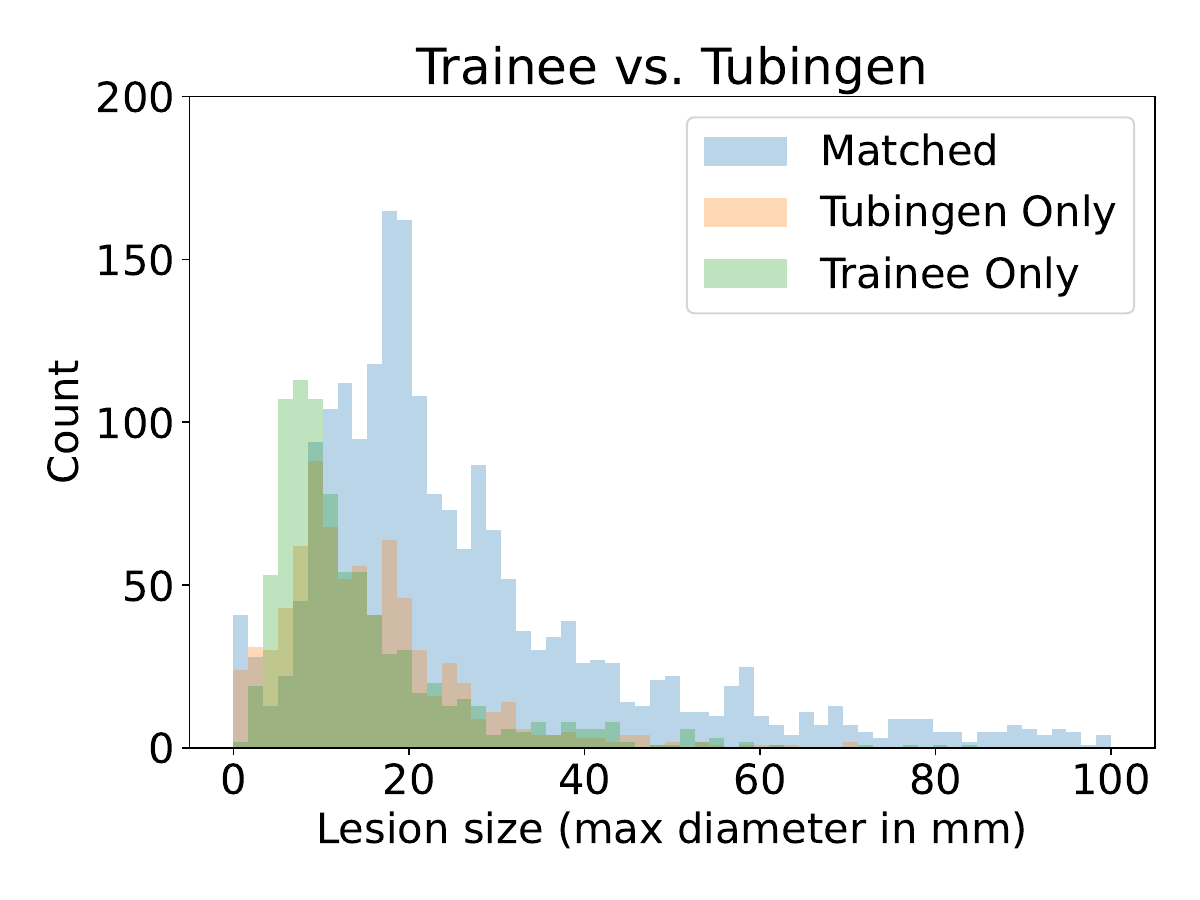}
  \caption{}
  \label{fig:sizecomp_trainee}
\end{subfigure}%
\begin{subfigure}{.33\textwidth}
  \centering
  \includegraphics[width=1\linewidth]{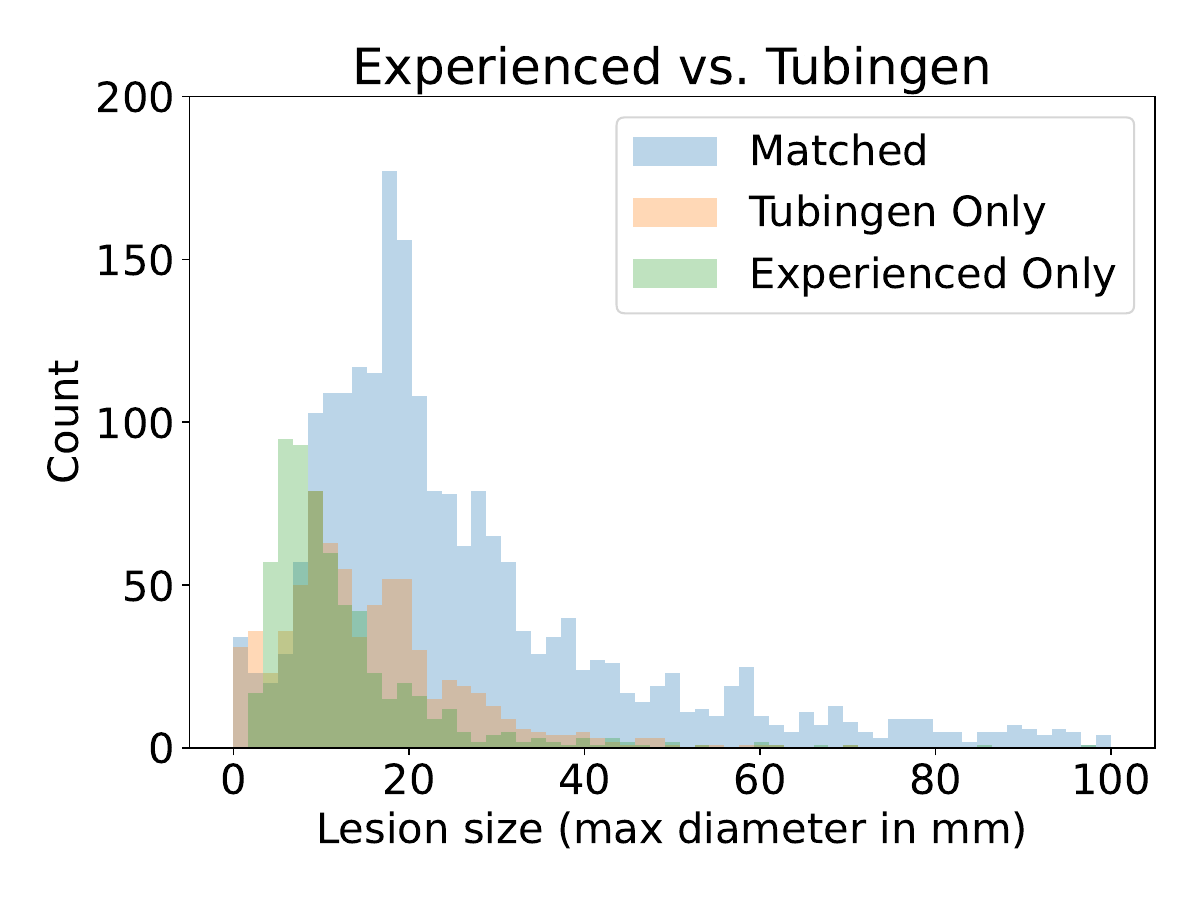}
  \caption{}
  \label{fig:sizecomp_exp}
\end{subfigure}
\caption{Inter-Reader Agreement across lesion sizes. }
\label{fig:lesion_level_analysis}
\end{figure}

Figure~\ref{fig:lesion_level_analysis} breaks down and visualizes the inter-reader agreement between the three sets of annotations by lesion sizes. Almost all disagreed lesions are small, with the majority smaller than $1~\mathrm{cm}$, which tend to be the most likely to be clinically indeterminate in routine practice, from a combination of co-existence of multiple reasonable differential diagnoses and physical limits of PET resolution (\textasciitilde5 mm)\textsuperscript{\cite{kudura2021malignancy,moses2011fundamental}}. In oncologic FDG-PET/CT, such findings frequently do not alter immediate management and are often followed longitudinally rather than definitively classified at a single time point. Therefore, reduced agreement in this subset reflects realistic diagnostic ambiguity rather than annotation error and highlights the importance of incorporating uncertainty into downstream modeling. The GazeXPErT lesion annotation also included an “uncertain”  versus “certain” tag for all accepted tumor lesions to account for this uncertainty in future model training. 

\subsection*{GazeXPErT Dataset Validation/Application Experiments}

\subsubsection*{Experiment 1: Value of expert attention for AI tumor segmentation models.}
We compared the performance of a standard 3D nnU-Net segmentation model\textsuperscript{\cite{isensee2021nnunet}} trained with and without incorporating gaze heatmap attentional information.

\textbf{Method}: The recommended nnU-Net ResEncL-Plan\textsuperscript{\cite{isensee2024nnunet}} was used for training with no major modification, apart from introducing the gaze volume as an additional input channel. 3D nnU-Net models were trained on the lesion pseudo-segmentation ground truth from the GazeXPErT dataset. Gaze heatmap volumes were created from raw gaze trajectories (see “3D processed volumes” under Data Records), with each voxel counting how often gaze trajectory intersects it after flattening out the temporal dimension. Gaze from experienced and trainee readers were combined (i.e. union) for input. Code to convert the processed gaze and pseudo-segmentation volumes into the folder structure required by nnU-Net (\texttt{create\_nnunet\_symlinks.py}) is included in our shared GitLab repository. 5-fold cross-validation was then performed directly via nnU-Net's command-line interface, with no additional custom training or evaluation code; fold assignment was handled internally over the pool of study reads, and only 3 of the 343 patients (<1\%) contributed two study reads to the dataset. The average Dice score is reported.

\textbf{Results}: The nnU-Net model that utilizes expert gaze attention heatmap performs consistently better across folds (Table~\ref{tab:nnunet}), despite our analysis showing that the readers spent < 8\% of the time actually looking at tumors. The readers spent far more time searching the remaining whole body for additional corroborating imaging clues. This is consistent with the hypothesis that there is information that can be distilled from experts' search patterns, which can be collected from radiologists' routine reading workflow. We note this result should be interpreted as a feasibility signal rather than strong evidence of a generalizable modeling gain: the segmentation ground truth used here was itself produced by the same class of human readers generating the gaze data, so shared human perceptual biases between the gaze signal and the reference labels may contribute to the observed improvement and cannot be fully disentangled from a genuine gaze-derived modeling benefit with the current experimental design.

\begin{table}[h!]
    \centering
    \begin{tabular}{l|c|c|c}
    \hline
     && \multicolumn{2}{c}{Lesion Level Metrics} \\
     & Dice Score & Precision &  Recall \\
     \hline
     \hline
     w/o Gaze  & $0.601 \pm 0.023$ & $0.506\pm 0.053$ & $0.774\pm 0.023$ \\
     with Gaze  & \textbf{0.682 $\pm$ 0.012} &  \textbf{0.665 $\pm$ 0.033} &  \textbf{0.857 $\pm$ 0.035} \\
     \hline
\end{tabular}
    \caption{3D nnU-Net model lesion segmentation performance with and without expert static gaze insight.}
    \label{tab:nnunet}
\end{table}

\begin{figure}[h!]
    \centering
    \includegraphics[width=\textwidth]{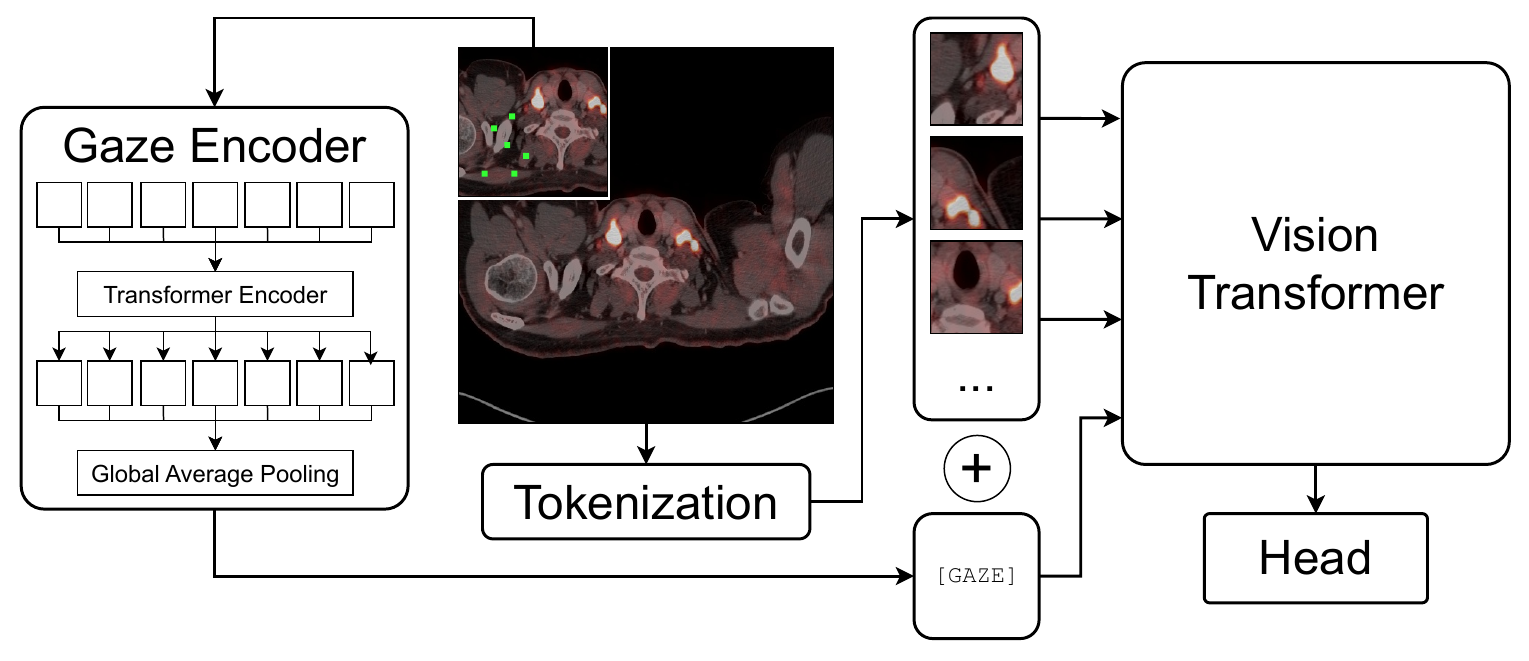}
    \caption{Transformer-based Gaze + Vision architecture. Gaze tokens derived from the trajectory are encoded into a \texttt{[GAZE]} token, then processed jointly by pretrained ViT.}
    \label{fig:gaze_transformer}
\end{figure}

\subsubsection*{Experiment 2: Last gaze point correction for improving lesion selection in a dynamic setting.}
The cost of eye-tracking devices can range widely (\$200--\$20,000 USD) depending on their licensing, setup, tracking rate (up to 500 Hz), accuracy and precision. Our motivation for this experiment is practical rather than diagnostic: reducing the offset between the last raw gaze point and the true lesion center can reduce spurious candidate proposal rejections during annotation (see Candidate Lesion Proposal Algorithm) and improve reader experience and throughput, particularly with more affordable, nonintrusive eye-tracking hardware. We do not claim that closing this gaze-to-lesion-center distance is itself a clinically meaningful endpoint; a reader can screen effectively for tumors without fixating precisely on their center. Rather, we show that modeling task-specific sequential gaze and image data can computationally correct the last raw gaze point to better approximate the target lesion location, without requiring higher-specification tracking hardware.

\textbf{Method}: We postprocess the GazeXPErT dataset, which initially derived 9,030 accepted gaze-to-lesion trajectories (See Data Records for COCO-style processed data). Invalid gaze samples were skipped and lesions with average gaze more than 100 pixels away were filtered, leaving n=8,610 lesion videos (6,947 train, 617 val, 1,046 test) suitable for this experiment. Train/val/test assignment followed the same patient-level splitting described in Data Records. An image-based Vision Transformer (VT) model pretrained on ImageNet\textsuperscript{\cite{deng2009imagenet}} ViT/B-32\textsuperscript{\cite{dosovitskiy2021image}} was extended by a Gaze Encoder module, and was jointly trained to predict a gaze point closer to the target lesion than the last gaze point (Figure~\ref{fig:gaze_transformer}). 

Specifically, the model inputs are up to 16 gaze coordinate samples before a lesion selection keystroke and the associated axial slice view image for the accepted lesion keystroke. We modified the VT’s CLS token and replaced it with learned gaze information. The Gaze Encoder in Figure~\ref{fig:gaze_transformer} is a normal transformer encoder, where tokenization is learned embedding and positional encoding is the same as in the original transformer paper\textsuperscript{\cite{vaswani2017attention}}. Loss is the average Hausdorff distance function between predicted point and all segmented pixels (as point coordinates), and is minimized when the prediction is in the center of a lesion.

\textbf{Results}: For evaluation, 10 randomly seeded runs on the test split (Table~\ref{tab:gaze_correction}) showed 74.95\% of the predicted gaze points are closer to the target lesions than the last gaze point. The results indicate that task-specific gaze correction modeling with temporal gaze and image search pattern data is able to lower the average gaze angle error for a dynamic targeting task from $1.134^\circ$ uncorrected to $0.856^\circ$ post correction, which is then within the expected range ($0.5^\circ$--$1^\circ$) for static targeting tasks. The correction also raised the fraction of predicted gaze points landing directly on the lesion mask from 6.11\% (raw last gaze) to 18.12\% (Table~\ref{tab:gaze_correction}), roughly a three-fold improvement. The absolute on-mask rate remains modest, consistent with the fact that human gaze naturally saccades around a region of interest rather than fixating precisely on it, but the relative improvement suggests the correction meaningfully sharpens gaze localization.

\begin{table}[h!]
    \centering
    \begin{tabular}{l|c|c|c|c}
    \hline
     & On Mask? ($\%$)& Improved Prediction? ($\%$)& Distance ($^\circ$) \\
     \hline
     \hline
     Raw Last Gaze  & 6.11\% & - & 1.134$^\circ$ \\
     Predicted Gaze  & \textbf{18.12\%} & \textbf{74.95\%} & \textbf{0.856$^\circ$}\\
     \hline
\end{tabular}
    \caption{Gaze correction experiment results. Average performance over 10 randomly initialized runs.}
    \label{tab:gaze_correction}
\end{table}

\subsubsection*{Experiment 3: Task-specific human expert intention prediction.}
Our final hypothesis is that temporal gaze data can help AI models detect \emph{when} a reader has committed to a specific lesion, rather than merely \emph{where} they are looking. Readers legitimately attend to non-tumor findings, supporting evidence, and incidental findings throughout routine interpretation, so the goal here is not to predict or restrict gaze location, but to detect the transition from open-ended visual search to an intentional, committed act of lesion selection. This is a potentially clinically useful and genuinely difficult task: target lesion detection and measurement is central to efficient FDG-PET/CT reads, yet readers spend most of their time searching for corroborating clues before committing to a target, making the search-to-intent transition non-trivial to predict. We therefore trained a model to test the feasibility of distinguishing “intentional” targeting from “unintentional” searching using gaze trajectories and their corresponding slice images.

\textbf{Method}: We modeled this problem as a binary classification task. Gaze trajectories and corresponding slice images at the moment of a reader’s deliberate lesion-directed keystroke are used as “intentional” samples (13,126 train, 1,391 val, 1,906 test). This class comprises every logged lesion-interaction intent, that is, selecting a candidate lesion, rejecting a candidate lesion, and subsequent adjustments to a candidate annotation such as resizing its bounding box, since the goal is to detect the act of committing to a specific target rather than the eventual accept/reject outcome. The 16,423 intentional samples therefore comprise the 9,030 accepted and 2,658 rejected lesion selections reported in Data Records, plus 4,735 other lesion interaction events. “Unintentional” samples are simulated from gaze trajectories and images sampled temporally away from any lesion-selection sequence, at random non-tumor PET hot spots (12,563 train, 1,209 val, 1,870 test). Train/val/test assignment followed the same patient-level splitting described in Data Records. We used the same vision transformer model architecture (see Figure~\ref{fig:gaze_transformer}) and similar input compared to experiment 2, with the difference being that the gaze trajectory used is longer (minimum 1 second, maximum 2 second at 60 Hz) and continuous, and that the model is trained with focal loss\textsuperscript{\cite{lin2017focal}}.

\textbf{Results}: Our results show that human intention prediction or understanding via gaze is not easy or a solved problem but is trainable, at least when posed as a classification problem (Table~\ref{tab:intent_classification}). We interpret this as a first proof-of-concept step toward gaze-triggered, non-intrusive AI assistance, for example cueing an interactive segmentation or measurement tool built on promptable segmentation foundation models\textsuperscript{\cite{kirillov2023segment}} (e.g. nnInteractive's 3D extension of this paradigm\textsuperscript{\cite{isensee2025nninteractive}}) at the moment a reader commits to a finding, rather than a system that directs or constrains reader attention. More advanced temporal modeling of sequential gaze and PET/CT images in future work will likely improve on these baseline results.

\begin{table}[h!]
    \centering
    \begin{tabular}{c|c|c|c}
    \hline
     Accuracy & Balanced (Macro) Accuracy & F1-Score & ROC AUC\\
     \hline
     \hline
     $67.53\%\pm 1.63$ & $67.51\%\pm 1.64$ & $0.673\pm 0.016$ & $0.747\pm 0.024$\\
     \hline
\end{tabular}
    \caption{Intention prediction experiment results. Average performance over 10 randomly initialized runs.}
    \label{tab:intent_classification}
\end{table}

\section*{Usage Notes}
\subsection*{Dataset Limitations}
There are a few key limitations in the GazeXPErT data annotation. Firstly, the PET/CT reading parameters were simplified to a single window for this proof-of-concept novel dataset. In real-life radiology read workflow, the human experts typically juggle between multiple viewing windows and screens, and have to compare multiple sequential PET/CT studies to arrive at the final report. However, we view the differences predominantly as software engineering challenges that future efforts with more resources can improve on.

Secondly, it is difficult to obtain pseudosegmentation ground truth for some lesion types with our current candidate proposal method. These include misregistered lesions (due to motion) and lesions that have similar FDG uptake relative to the background or have indistinct margins on PET and CT. Lesions with heterogeneous FDG uptake can also be selected as multiple lesions instead of one larger lesion. These complex lesions constitute only a small portion of the GazeXPErT dataset, since the included cancer types are relatively hypermetabolic and most of the Tübingen FDG-PET/CT studies used are baseline staging or first post therapy follow up studies. However, there are potential challenges with our methodology for other less hypermetabolic cancers (e.g. thyroid cancer) or cancers that occur in organs that have a high background FDG uptake (e.g. renal cancers). 

In addition, lesion conspicuity and reader search behavior may be influenced by variability in PET/CT acquisition parameters, including scanner type, reconstruction method, uptake time, and patient preparation. Although such variability reflects real-world clinical practice, it can introduce heterogeneity in gaze patterns that should be considered when training and validating downstream models.

The gaze data were collected using a single screen-based eye-tracking device (Tobii Pro Spark) and a single custom annotation platform; portability of our findings to other eye-tracking hardware, sampling rates, display configurations, or annotation software has not been evaluated, and gaze behavior could differ under a different capture setup. The recorded reading times likewise reflect this simplified, single-viewer proof-of-concept setup and should not be interpreted as representative of reading times in ordinary multi-monitor clinical workflows, which we did not measure here.

As described in the Methods, our select-then-accept/reject annotation design was intended to mark a window of unequivocal recognition rather than relying on gaze dwell alone. Subtler perceptual phenomena, such as covert attention or momentary uncertainty before a reader commits to a selection, cannot be fully ruled out.

\subsection*{Baseline Gaze Post-processing and Modeling Limitations}
For our validation experiments, we used fairly simple, established gaze-processing methods and well-known AI model architectures, without significant task-specific alteration or gaze-specific pretraining. This work does not characterize model generalizability across reader experience level, institution, or acquisition platform, given the modest number of readers in this proof-of-concept dataset; we see rigorous generalizability testing as a natural next step for studies built on this released dataset. A separate, unresolved technical question is how best to assign the relevant window of a continuous gaze trajectory to a given lesion. Because we wanted to minimize alteration to radiologists' natural search patterns, readers could scroll freely back and forth through the volume, so a lesion may be visited multiple times with no guarantee that a reader's attention is on it during every adjacent-slice visit. We see more sophisticated gaze processing and modeling — for example, a pretrained, task-agnostic gaze model that better leverages radiologists' temporal search patterns — as a key direction for future work on interactive agents that understand expert intent through gaze.

\subsection*{Conclusion and Future Work}
We have demonstrated basic gaze postprocessing techniques that make complex raw gaze data useful for a subset of baseline tasks: 3D volumetric lesion segmentation with expert insight, and multimodal temporal gaze prediction for last-gaze-point correction and task-intention prediction. As detailed above, these are feasibility demonstrations of the kinds of tasks GazeXPErT can support, representing only a subset of the problems the dataset is intended to enable. Additional future directions include visual grounding, causal reasoning, and clinically relevant imaging feature augmentation for FDG-PET/CT, as well as more advanced human-computer interaction, human intention prediction, expert gaze-rewarded modeling, and gaze-prompted interactive segmentation approaches such as nnInteractive\textsuperscript{\cite{isensee2025nninteractive}}.

From a clinical workflow perspective, datasets that capture expert visual reasoning may help enable AI systems that augment, rather than replace, physician interpretation\textsuperscript{\cite{khosravan2019collaborative,saab2021observational,droguet2024automated,frood2024comparative}}. We envision gaze as a natural complement to today's keyboard-, mouse-, and voice-based interfaces: a faster, more dynamic channel for AI-human collaboration that could support broader clinical adoption — for example, gaze-aware models that flag overlooked regions or adapt to inferred reader intent. Realizing this vision, and empirically establishing whether it improves efficiency, transparency, or cognitive burden in oncologic imaging, is a direction for future work building on the released dataset.

From the dataset perspective, future work includes expanding to other cancer types, as well as testing the data collection methodology for PET/CTs using non-FDG radiotracers, such as Prostate-Specific Membrane Antigen (PSMA) tracers for prostate cancer.
  
Overall, GazeXPErT is a dataset packed with reasoning search patterns from human experts collected in a simulated routine FDG-PET/CT read workflow. In contributing this dataset, our goal is not only to aid development of more interpretable, explainable and robust lesion detection or segmentation models, but also to motivate the community toward a new generation of interactive AI that learns continuously from radiologists' gaze-informed decisions during daily practice — closing a feedback loop in which the AI and the human reader improve together over time, rather than the AI remaining a static tool retrained only offline. This is a future research direction that GazeXPErT is intended to help seed, not a capability demonstrated by the present work.

\section*{Code availability}
Code and detailed software version information for the gaze data postprocessing and the validation experiments is available at: \url{https://zivgitlab.uni-muenster.de/cvmls/gazexpert}


\bibliography{eye_track}

\begin{thebibliography}{10}
\urlstyle{rm}
\expandafter\ifx\csname url\endcsname\relax
  \def\url#1{\texttt{#1}}\fi
\expandafter\ifx\csname urlprefix\endcsname\relax\def\urlprefix{URL }\fi
\expandafter\ifx\csname doiprefix\endcsname\relax\def\doiprefix{DOI: }\fi
\providecommand{\bibinfo}[2]{#2}
\providecommand{\eprint}[2][]{\url{#2}}

\bibitem{parihar2023fdg}
\bibinfo{author}{Parihar, A.~S.}, \bibinfo{author}{Dehdashti, F.} \&
  \bibinfo{author}{Wahl, R.~L.}
\newblock \bibinfo{journal}{\bibinfo{title}{Fdg pet/ct--based response
  assessment in malignancies}}.
\newblock {\emph{\JournalTitle{Radiographics}}} \textbf{\bibinfo{volume}{43}},
  \bibinfo{pages}{e220122} (\bibinfo{year}{2023}).

\bibitem{wahl2009recist}
\bibinfo{author}{Wahl, R.~L.}, \bibinfo{author}{Jacene, H.},
  \bibinfo{author}{Kasamon, Y.} \& \bibinfo{author}{Lodge, M.~A.}
\newblock \bibinfo{journal}{\bibinfo{title}{From recist to percist: evolving
  considerations for pet response criteria in solid tumors}}.
\newblock {\emph{\JournalTitle{Journal of nuclear medicine}}}
  \textbf{\bibinfo{volume}{50}}, \bibinfo{pages}{122S--150S}
  (\bibinfo{year}{2009}).

\bibitem{jadvar2017appropriate}
\bibinfo{author}{Jadvar, H.} \emph{et~al.}
\newblock \bibinfo{journal}{\bibinfo{title}{Appropriate use criteria for
  18f-fdg pet/ct in restaging and treatment response assessment of malignant
  disease}}.
\newblock {\emph{\JournalTitle{Journal of Nuclear Medicine}}}
  \textbf{\bibinfo{volume}{58}}, \bibinfo{pages}{2026--2037}
  (\bibinfo{year}{2017}).

\bibitem{czernin2002clinical}
\bibinfo{author}{Czernin, J.}
\newblock \bibinfo{journal}{\bibinfo{title}{Clinical applications of fdg-pet in
  oncology}}.
\newblock {\emph{\JournalTitle{Acta Medica Austriaca}}}
  \textbf{\bibinfo{volume}{29}}, \bibinfo{pages}{162--170}
  (\bibinfo{year}{2002}).

\bibitem{weber2005use}
\bibinfo{author}{Weber, W.~A.}
\newblock \bibinfo{journal}{\bibinfo{title}{Use of pet for monitoring cancer
  therapy and for predicting outcome}}.
\newblock {\emph{\JournalTitle{Journal of Nuclear Medicine}}}
  \textbf{\bibinfo{volume}{46}}, \bibinfo{pages}{983--995}
  (\bibinfo{year}{2005}).

\bibitem{lardinois2003staging}
\bibinfo{author}{Lardinois, D.} \emph{et~al.}
\newblock \bibinfo{journal}{\bibinfo{title}{Staging of non--small-cell lung
  cancer with integrated positron-emission tomography and computed
  tomography}}.
\newblock {\emph{\JournalTitle{New England Journal of Medicine}}}
  \textbf{\bibinfo{volume}{348}}, \bibinfo{pages}{2500--2507}
  (\bibinfo{year}{2003}).

\bibitem{boellaard2015fdg}
\bibinfo{author}{Boellaard, R.} \emph{et~al.}
\newblock \bibinfo{journal}{\bibinfo{title}{Fdg pet/ct: Eanm procedure
  guidelines for tumour imaging: version 2.0}}.
\newblock {\emph{\JournalTitle{European journal of nuclear medicine and
  molecular imaging}}} \textbf{\bibinfo{volume}{42}}, \bibinfo{pages}{328--354}
  (\bibinfo{year}{2015}).

\bibitem{avril2005monitoring}
\bibinfo{author}{Avril, N.~E.} \& \bibinfo{author}{Weber, W.~A.}
\newblock \bibinfo{journal}{\bibinfo{title}{Monitoring response to treatment in
  patients utilizing pet}}.
\newblock {\emph{\JournalTitle{Radiologic Clinics}}}
  \textbf{\bibinfo{volume}{43}}, \bibinfo{pages}{189--204}
  (\bibinfo{year}{2005}).

\bibitem{fletcher2008recommendations}
\bibinfo{author}{Fletcher, J.~W.} \emph{et~al.}
\newblock \bibinfo{journal}{\bibinfo{title}{Recommendations on the use of
  18f-fdg pet in oncology}}.
\newblock {\emph{\JournalTitle{Journal of Nuclear Medicine}}}
  \textbf{\bibinfo{volume}{49}}, \bibinfo{pages}{480--508}
  (\bibinfo{year}{2008}).

\bibitem{delgado2003meta}
\bibinfo{author}{Delgado-Bolton, R.~C.},
  \bibinfo{author}{Fern{\'a}ndez-P{\'e}rez, C.},
  \bibinfo{author}{Gonz{\'a}lez-Mat{\'e}, A.} \& \bibinfo{author}{Carreras,
  J.~L.}
\newblock \bibinfo{journal}{\bibinfo{title}{Meta-analysis of the performance of
  18f-fdg pet in primary tumor detection in unknown primary tumors}}.
\newblock {\emph{\JournalTitle{Journal of Nuclear Medicine}}}
  \textbf{\bibinfo{volume}{44}}, \bibinfo{pages}{1301--1314}
  (\bibinfo{year}{2003}).

\bibitem{gregoire2010pet}
\bibinfo{author}{Gregoire, V.} \& \bibinfo{author}{Chiti, A.}
\newblock \bibinfo{journal}{\bibinfo{title}{Pet in radiotherapy planning:
  particularly exquisite test or pending and experimental tool?}}
\newblock {\emph{\JournalTitle{Radiotherapy and Oncology}}}
  \textbf{\bibinfo{volume}{96}}, \bibinfo{pages}{275--276}
  (\bibinfo{year}{2010}).

\bibitem{thorwarth2012integration}
\bibinfo{author}{Thorwarth, D.} \emph{et~al.}
\newblock \bibinfo{journal}{\bibinfo{title}{Integration of fdg-pet/ct into
  external beam radiation therapy planning}}.
\newblock {\emph{\JournalTitle{Nuklearmedizin-NuclearMedicine}}}
  \textbf{\bibinfo{volume}{51}}, \bibinfo{pages}{140--153}
  (\bibinfo{year}{2012}).

\bibitem{hofman2016we}
\bibinfo{author}{Hofman, M.~S.} \& \bibinfo{author}{Hicks, R.~J.}
\newblock \bibinfo{journal}{\bibinfo{title}{How we read oncologic fdg pet/ct}}.
\newblock {\emph{\JournalTitle{Cancer Imaging}}} \textbf{\bibinfo{volume}{16}},
  \bibinfo{pages}{35} (\bibinfo{year}{2016}).

\bibitem{zhang2013prognostic}
\bibinfo{author}{Zhang, H.} \emph{et~al.}
\newblock \bibinfo{journal}{\bibinfo{title}{Prognostic value of metabolic tumor
  burden from 18f-fdg pet in surgical patients with non--small-cell lung
  cancer}}.
\newblock {\emph{\JournalTitle{Academic radiology}}}
  \textbf{\bibinfo{volume}{20}}, \bibinfo{pages}{32--40}
  (\bibinfo{year}{2013}).

\bibitem{meignan2021total}
\bibinfo{author}{Meignan, M.}, \bibinfo{author}{Cottereau, A.-S.},
  \bibinfo{author}{Specht, L.} \& \bibinfo{author}{Mikhaeel, N.~G.}
\newblock \bibinfo{journal}{\bibinfo{title}{Total tumor burden in lymphoma--an
  evolving strong prognostic parameter}}.
\newblock {\emph{\JournalTitle{The British journal of radiology}}}
  \textbf{\bibinfo{volume}{94}}, \bibinfo{pages}{20210448}
  (\bibinfo{year}{2021}).

\bibitem{christensen2025projected}
\bibinfo{author}{Christensen, E.~W.}, \bibinfo{author}{Parikh, J.~R.},
  \bibinfo{author}{Drake, A.~R.}, \bibinfo{author}{Rubin, E.~M.} \&
  \bibinfo{author}{Rula, E.~Y.}
\newblock \bibinfo{journal}{\bibinfo{title}{Projected us radiologist supply,
  2025 to 2055}}.
\newblock {\emph{\JournalTitle{Journal of the American College of Radiology}}}
  \textbf{\bibinfo{volume}{22}}, \bibinfo{pages}{161--169}
  (\bibinfo{year}{2025}).

\bibitem{gatidis2024results}
\bibinfo{author}{Gatidis, S.} \emph{et~al.}
\newblock \bibinfo{journal}{\bibinfo{title}{Results from the autopet challenge
  on fully automated lesion segmentation in oncologic pet/ct imaging}}.
\newblock {\emph{\JournalTitle{Nature Machine Intelligence}}}
  \textbf{\bibinfo{volume}{6}}, \bibinfo{pages}{1396--1405}
  (\bibinfo{year}{2024}).

\bibitem{grossiord2017automated}
\bibinfo{author}{Grossiord, E.}, \bibinfo{author}{Talbot, H.},
  \bibinfo{author}{Passat, N.}, \bibinfo{author}{Meignan, M.} \&
  \bibinfo{author}{Najman, L.}
\newblock \bibinfo{title}{Automated 3d lymphoma lesion segmentation from pet/ct
  characteristics}.
\newblock In \emph{\bibinfo{booktitle}{2017 IEEE 14th international symposium
  on biomedical imaging (ISBI 2017)}}, \bibinfo{pages}{174--178}
  (\bibinfo{organization}{IEEE}, \bibinfo{year}{2017}).

\bibitem{eyuboglu2021multi}
\bibinfo{author}{Eyuboglu, S.} \emph{et~al.}
\newblock \bibinfo{journal}{\bibinfo{title}{Multi-task weak supervision enables
  anatomically-resolved abnormality detection in whole-body fdg-pet/ct}}.
\newblock {\emph{\JournalTitle{Nature communications}}}
  \textbf{\bibinfo{volume}{12}}, \bibinfo{pages}{1880} (\bibinfo{year}{2021}).

\bibitem{gatidis2022whole}
\bibinfo{author}{Gatidis, S.} \emph{et~al.}
\newblock \bibinfo{journal}{\bibinfo{title}{A whole-body fdg-pet/ct dataset
  with manually annotated tumor lesions}}.
\newblock {\emph{\JournalTitle{Scientific Data}}} \textbf{\bibinfo{volume}{9}},
  \bibinfo{pages}{601}, \url{10.1038/s41597-022-01718-3}
  (\bibinfo{year}{2022}).

\bibitem{xia2025comprehensive}
\bibinfo{author}{Xia, Q.} \emph{et~al.}
\newblock \bibinfo{journal}{\bibinfo{title}{A comprehensive review of deep
  learning for medical image segmentation}}.
\newblock {\emph{\JournalTitle{Neurocomputing}}}
  \textbf{\bibinfo{volume}{613}}, \bibinfo{pages}{128740}
  (\bibinfo{year}{2025}).

\bibitem{salimi2024explainable}
\bibinfo{author}{Salimi, Y.}, \bibinfo{author}{Mansouri, Z.},
  \bibinfo{author}{Amini, M.}, \bibinfo{author}{Mainta, I.} \&
  \bibinfo{author}{Zaidi, H.}
\newblock \bibinfo{journal}{\bibinfo{title}{Explainable ai for automated
  respiratory misalignment detection in pet/ct imaging}}.
\newblock {\emph{\JournalTitle{Physics in Medicine \& Biology}}}
  \textbf{\bibinfo{volume}{69}}, \bibinfo{pages}{215036}
  (\bibinfo{year}{2024}).

\bibitem{mbakwe2023fairness}
\bibinfo{author}{Mbakwe, A.~B.}, \bibinfo{author}{Lourentzou, I.},
  \bibinfo{author}{Celi, L.~A.} \& \bibinfo{author}{Wu, J.~T.}
\newblock \bibinfo{journal}{\bibinfo{title}{Fairness metrics for health ai: we
  have a long way to go}}.
\newblock {\emph{\JournalTitle{EBioMedicine}}} \textbf{\bibinfo{volume}{90}}
  (\bibinfo{year}{2023}).

\bibitem{sundar2024automatic}
\bibinfo{author}{Sundar, L. K.~S.} \& \bibinfo{author}{Beyer, T.}
\newblock \bibinfo{journal}{\bibinfo{title}{Is automatic tumor segmentation on
  whole-body 18f-fdg pet images a clinical reality?}}
\newblock {\emph{\JournalTitle{Journal of Nuclear Medicine}}}
  \textbf{\bibinfo{volume}{65}}, \bibinfo{pages}{995--997}
  (\bibinfo{year}{2024}).

\bibitem{recht2020integrating}
\bibinfo{author}{Recht, M.~P.} \emph{et~al.}
\newblock \bibinfo{journal}{\bibinfo{title}{Integrating artificial intelligence
  into the clinical practice of radiology: challenges and recommendations}}.
\newblock {\emph{\JournalTitle{European Radiology}}}
  \textbf{\bibinfo{volume}{30}}, \bibinfo{pages}{3576--3584},
  \url{10.1007/s00330-020-06672-5} (\bibinfo{year}{2020}).

\bibitem{droguet2024automated}
\bibinfo{author}{Droguet, M.} \emph{et~al.}
\newblock \bibinfo{title}{Automated segmentation of lesions in [18f] fdg pet/ct
  images of lung cancer patients: external evaluation of an ai-driven lesion
  segmentation tool (lion)} (\bibinfo{year}{2024}).

\bibitem{frood2024comparative}
\bibinfo{author}{Frood, R.} \emph{et~al.}
\newblock \bibinfo{journal}{\bibinfo{title}{Comparative effectiveness of
  standard vs. ai-assisted pet/ct reading workflow for pre-treatment lymphoma
  staging: a multi-institutional reader study evaluation}}.
\newblock {\emph{\JournalTitle{Frontiers in Nuclear Medicine}}}
  \textbf{\bibinfo{volume}{3}}, \bibinfo{pages}{1327186}
  (\bibinfo{year}{2024}).

\bibitem{khosravan2019collaborative}
\bibinfo{author}{Khosravan, N.} \emph{et~al.}
\newblock \bibinfo{journal}{\bibinfo{title}{A collaborative computer aided
  diagnosis (c-cad) system with eye-tracking, sparse attentional model, and
  deep learning}}.
\newblock {\emph{\JournalTitle{Medical image analysis}}}
  \textbf{\bibinfo{volume}{51}}, \bibinfo{pages}{101--115}
  (\bibinfo{year}{2019}).

\bibitem{hayhoe2014modeling}
\bibinfo{author}{Hayhoe, M.} \& \bibinfo{author}{Ballard, D.}
\newblock \bibinfo{journal}{\bibinfo{title}{Modeling task control of eye
  movements}}.
\newblock {\emph{\JournalTitle{Current Biology}}}
  \textbf{\bibinfo{volume}{24}}, \bibinfo{pages}{R622--R628}
  (\bibinfo{year}{2014}).

\bibitem{foulsham2015eye}
\bibinfo{author}{Foulsham, T.}
\newblock \bibinfo{journal}{\bibinfo{title}{Eye movements and their functions
  in everyday tasks}}.
\newblock {\emph{\JournalTitle{Eye}}} \textbf{\bibinfo{volume}{29}},
  \bibinfo{pages}{196--199} (\bibinfo{year}{2015}).

\bibitem{van2017visual}
\bibinfo{author}{Van~der Gijp, A.} \emph{et~al.}
\newblock \bibinfo{journal}{\bibinfo{title}{How visual search relates to visual
  diagnostic performance: a narrative systematic review of eye-tracking
  research in radiology}}.
\newblock {\emph{\JournalTitle{Advances in Health Sciences Education}}}
  \textbf{\bibinfo{volume}{22}}, \bibinfo{pages}{765--787}
  (\bibinfo{year}{2017}).

\bibitem{waite2019analysis}
\bibinfo{author}{Waite, S.} \emph{et~al.}
\newblock \bibinfo{journal}{\bibinfo{title}{Analysis of perceptual expertise in
  radiology--current knowledge and a new perspective}}.
\newblock {\emph{\JournalTitle{Frontiers in human neuroscience}}}
  \textbf{\bibinfo{volume}{13}}, \bibinfo{pages}{213} (\bibinfo{year}{2019}).

\bibitem{mall2018modeling}
\bibinfo{author}{Mall, S.}, \bibinfo{author}{Brennan, P.~C.} \&
  \bibinfo{author}{Mello-Thoms, C.}
\newblock \bibinfo{journal}{\bibinfo{title}{Modeling visual search behavior of
  breast radiologists using a deep convolution neural network}}.
\newblock {\emph{\JournalTitle{Journal of medical imaging}}}
  \textbf{\bibinfo{volume}{5}}, \bibinfo{pages}{035502--035502}
  (\bibinfo{year}{2018}).

\bibitem{karargyris2021creation}
\bibinfo{author}{Karargyris, A.} \emph{et~al.}
\newblock \bibinfo{journal}{\bibinfo{title}{Creation and validation of a chest
  x-ray dataset with eye-tracking and report dictation for ai development}}.
\newblock {\emph{\JournalTitle{Scientific data}}} \textbf{\bibinfo{volume}{8}},
  \bibinfo{pages}{92} (\bibinfo{year}{2021}).

\bibitem{saab2021observational}
\bibinfo{author}{Saab, K.} \emph{et~al.}
\newblock \bibinfo{title}{Observational supervision for medical image
  classification using gaze data}.
\newblock In \emph{\bibinfo{booktitle}{International conference on medical
  image computing and computer-assisted Intervention}},
  \bibinfo{pages}{603--614} (\bibinfo{organization}{Springer},
  \bibinfo{year}{2021}).

\bibitem{wang2025improving}
\bibinfo{author}{Wang, S.} \emph{et~al.}
\newblock \bibinfo{journal}{\bibinfo{title}{Improving self-supervised medical
  image pre-training by early alignment with human eye gaze information}}.
\newblock {\emph{\JournalTitle{IEEE Transactions on Medical Imaging}}}
  (\bibinfo{year}{2025}).

\bibitem{zhou2021using}
\bibinfo{author}{Zhou, F.}, \bibinfo{author}{Yang, X.~J.} \&
  \bibinfo{author}{De~Winter, J.~C.}
\newblock \bibinfo{journal}{\bibinfo{title}{Using eye-tracking data to predict
  situation awareness in real time during takeover transitions in conditionally
  automated driving}}.
\newblock {\emph{\JournalTitle{IEEE transactions on intelligent transportation
  systems}}} \textbf{\bibinfo{volume}{23}}, \bibinfo{pages}{2284--2295}
  (\bibinfo{year}{2021}).

\bibitem{cho2024integration}
\bibinfo{author}{Cho, S.-W.}, \bibinfo{author}{Lim, Y.-H.},
  \bibinfo{author}{Seo, K.-M.} \& \bibinfo{author}{Kim, J.}
\newblock \bibinfo{journal}{\bibinfo{title}{Integration of eye-tracking and
  object detection in a deep learning system for quality inspection analysis}}.
\newblock {\emph{\JournalTitle{Journal of Computational Design and
  Engineering}}} \textbf{\bibinfo{volume}{11}}, \bibinfo{pages}{158--173}
  (\bibinfo{year}{2024}).

\bibitem{zhang2022can}
\bibinfo{author}{Zhang, Z.}, \bibinfo{author}{Crandall, D.},
  \bibinfo{author}{Proulx, M.}, \bibinfo{author}{Talathi, S.} \&
  \bibinfo{author}{Sharma, A.}
\newblock \bibinfo{title}{Can gaze inform egocentric action recognition?}
\newblock In \emph{\bibinfo{booktitle}{2022 Symposium on Eye Tracking Research
  and Applications}}, \bibinfo{pages}{1--7} (\bibinfo{year}{2022}).

\bibitem{hessels2019eye}
\bibinfo{author}{Hessels, R.~S.} \& \bibinfo{author}{Hooge, I.~T.}
\newblock \bibinfo{journal}{\bibinfo{title}{Eye tracking in developmental
  cognitive neuroscience--the good, the bad and the ugly}}.
\newblock {\emph{\JournalTitle{Developmental cognitive neuroscience}}}
  \textbf{\bibinfo{volume}{40}}, \bibinfo{pages}{100710}
  (\bibinfo{year}{2019}).

\bibitem{rizzo2022machine}
\bibinfo{author}{Rizzo, A.}, \bibinfo{author}{Ermini, S.},
  \bibinfo{author}{Zanca, D.}, \bibinfo{author}{Bernabini, D.} \&
  \bibinfo{author}{Rossi, A.}
\newblock \bibinfo{journal}{\bibinfo{title}{A machine learning approach for
  detecting cognitive interference based on eye-tracking data}}.
\newblock {\emph{\JournalTitle{Frontiers in Human Neuroscience}}}
  \textbf{\bibinfo{volume}{16}}, \bibinfo{pages}{806330}
  (\bibinfo{year}{2022}).

\bibitem{duan2025eye}
\bibinfo{author}{Duan, J.}, \bibinfo{author}{Zhang, M.}, \bibinfo{author}{Song,
  M.}, \bibinfo{author}{Xu, X.} \& \bibinfo{author}{Lu, H.}
\newblock \bibinfo{journal}{\bibinfo{title}{Eye tracking-enhanced deep learning
  for medical image analysis: A systematic review on data efficiency,
  interpretability, and multimodal integration}}.
\newblock {\emph{\JournalTitle{Bioengineering}}} \textbf{\bibinfo{volume}{12}},
  \bibinfo{pages}{954} (\bibinfo{year}{2025}).

\bibitem{moradizeyveh2024eye}
\bibinfo{author}{Moradizeyveh, S.} \emph{et~al.}
\newblock \bibinfo{journal}{\bibinfo{title}{When eye-tracking meets machine
  learning: A systematic review on applications in medical image analysis}}.
\newblock {\emph{\JournalTitle{arXiv preprint arXiv:2403.07834}}}
  (\bibinfo{year}{2024}).

\bibitem{bigolin2022reflacx}
\bibinfo{author}{Bigolin~Lanfredi, R.} \emph{et~al.}
\newblock \bibinfo{journal}{\bibinfo{title}{Reflacx, a dataset of reports and
  eye-tracking data for localization of abnormalities in chest x-rays}}.
\newblock {\emph{\JournalTitle{Scientific data}}} \textbf{\bibinfo{volume}{9}},
  \bibinfo{pages}{350} (\bibinfo{year}{2022}).

\bibitem{pham2024fg}
\bibinfo{author}{Pham, T.~T.} \emph{et~al.}
\newblock \bibinfo{title}{Fg-cxr: a radiologist-aligned gaze dataset for
  enhancing interpretability in chest x-ray report generation}.
\newblock In \emph{\bibinfo{booktitle}{Proceedings of the Asian conference on
  computer vision}}, \bibinfo{pages}{941--958} (\bibinfo{year}{2024}).

\bibitem{ma2024eye}
\bibinfo{author}{Ma, C.} \emph{et~al.}
\newblock \bibinfo{journal}{\bibinfo{title}{Eye-gaze guided multi-modal
  alignment for medical representation learning}}.
\newblock {\emph{\JournalTitle{Advances in Neural Information Processing
  Systems}}} \textbf{\bibinfo{volume}{37}}, \bibinfo{pages}{6126--6153}
  (\bibinfo{year}{2024}).

\bibitem{wang2024gazegnn}
\bibinfo{author}{Wang, B.} \emph{et~al.}
\newblock \bibinfo{title}{Gazegnn: A gaze-guided graph neural network for chest
  x-ray classification}.
\newblock In \emph{\bibinfo{booktitle}{Proceedings of the IEEE/CVF Winter
  Conference on Applications of Computer Vision}}, \bibinfo{pages}{2194--2203}
  (\bibinfo{year}{2024}).

\bibitem{zhu2022gaze}
\bibinfo{author}{Zhu, H.}, \bibinfo{author}{Salcudean, S.} \&
  \bibinfo{author}{Rohling, R.}
\newblock \bibinfo{title}{Gaze-guided class activation mapping: Leverage human
  visual attention for network attention in chest x-rays classification}.
\newblock In \emph{\bibinfo{booktitle}{Proceedings of the 15th International
  Symposium on Visual Information Communication and Interaction}},
  \bibinfo{pages}{1--8} (\bibinfo{year}{2022}).

\bibitem{hsieh2024eyexnet}
\bibinfo{author}{Hsieh, C.} \emph{et~al.}
\newblock \bibinfo{journal}{\bibinfo{title}{{EyeXNet}: Enhancing abnormality
  detection and diagnosis via eye-tracking and x-ray fusion}}.
\newblock {\emph{\JournalTitle{Machine Learning and Knowledge Extraction}}}
  \textbf{\bibinfo{volume}{6}}, \bibinfo{pages}{1055--1071},
  \url{10.3390/make6020048} (\bibinfo{year}{2024}).

\bibitem{khaertdinova2024gaze}
\bibinfo{author}{Khaertdinova, L.}, \bibinfo{author}{Pershin, I.},
  \bibinfo{author}{Shmykova, T.} \& \bibinfo{author}{Ibragimov, B.}
\newblock \bibinfo{title}{Gaze-assisted medical image segmentation}.
\newblock In \emph{\bibinfo{booktitle}{NeurIPS 2024 Workshop on Advancements In
  Medical Foundation Models: Explainability, Robustness, Security, and Beyond
  (AIM-FM)}} (\bibinfo{year}{2024}).
\newblock \bibinfo{note}{ArXiv:2410.17920}.

\bibitem{stember2019eye}
\bibinfo{author}{Stember, J.~N.} \emph{et~al.}
\newblock \bibinfo{journal}{\bibinfo{title}{Eye tracking for deep learning
  segmentation using convolutional neural networks}}.
\newblock {\emph{\JournalTitle{Journal of digital imaging}}}
  \textbf{\bibinfo{volume}{32}}, \bibinfo{pages}{597--604}
  (\bibinfo{year}{2019}).

\bibitem{khosravan2016gaze2segment}
\bibinfo{author}{Khosravan, N.} \emph{et~al.}
\newblock \bibinfo{title}{Gaze2segment: a pilot study for integrating
  eye-tracking technology into medical image segmentation}.
\newblock In \emph{\bibinfo{booktitle}{Bayesian and grAphical Models for
  Biomedical Imaging}}, \bibinfo{pages}{94--104}
  (\bibinfo{publisher}{Springer}, \bibinfo{year}{2016}).

\bibitem{kirillov2023segment}
\bibinfo{author}{Kirillov, A.} \emph{et~al.}
\newblock \bibinfo{title}{Segment anything}.
\newblock In \emph{\bibinfo{booktitle}{Proceedings of the IEEE/CVF
  International Conference on Computer Vision}}, \bibinfo{pages}{4015--4026}
  (\bibinfo{year}{2023}).

\bibitem{ma2024segment}
\bibinfo{author}{Ma, J.} \emph{et~al.}
\newblock \bibinfo{journal}{\bibinfo{title}{Segment anything in medical
  images}}.
\newblock {\emph{\JournalTitle{Nature Communications}}}
  \textbf{\bibinfo{volume}{15}}, \bibinfo{pages}{654},
  \url{10.1038/s41467-024-44824-z} (\bibinfo{year}{2024}).

\bibitem{pham2025ct}
\bibinfo{author}{Pham, T.~T.} \emph{et~al.}
\newblock \bibinfo{title}{Ct-scangaze: A dataset and baselines for 3d
  volumetric scanpath modeling}.
\newblock In \emph{\bibinfo{booktitle}{Proceedings of the IEEE/CVF
  International Conference on Computer Vision}}, \bibinfo{pages}{21732--21743}
  (\bibinfo{year}{2025}).

\bibitem{rostianingsih2020coco}
\bibinfo{author}{Rostianingsih, S.}, \bibinfo{author}{Setiawan, A.} \&
  \bibinfo{author}{Halim, C.~I.}
\newblock \bibinfo{journal}{\bibinfo{title}{Coco (creating common object in
  context) dataset for chemistry apparatus}}.
\newblock {\emph{\JournalTitle{Procedia Computer Science}}}
  \textbf{\bibinfo{volume}{171}}, \bibinfo{pages}{2445--2452}
  (\bibinfo{year}{2020}).

\bibitem{kundel1978visual}
\bibinfo{author}{Kundel, H.~L.}, \bibinfo{author}{Nodine, C.~F.} \&
  \bibinfo{author}{Carmody, D.}
\newblock \bibinfo{journal}{\bibinfo{title}{Visual scanning, pattern
  recognition and decision-making in pulmonary nodule detection}}.
\newblock {\emph{\JournalTitle{Investigative Radiology}}}
  \textbf{\bibinfo{volume}{13}}, \bibinfo{pages}{175--181}
  (\bibinfo{year}{1978}).

\bibitem{krupinski2010current}
\bibinfo{author}{Krupinski, E.~A.}
\newblock \bibinfo{journal}{\bibinfo{title}{Current perspectives in medical
  image perception}}.
\newblock {\emph{\JournalTitle{Attention, Perception, \& Psychophysics}}}
  \textbf{\bibinfo{volume}{72}}, \bibinfo{pages}{1205--1217},
  \url{10.3758/APP.72.5.1205} (\bibinfo{year}{2010}).

\bibitem{drew2013invisible}
\bibinfo{author}{Drew, T.}, \bibinfo{author}{V{\~o}, M. L.-H.} \&
  \bibinfo{author}{Wolfe, J.~M.}
\newblock \bibinfo{journal}{\bibinfo{title}{The invisible gorilla strikes
  again: Sustained inattentional blindness in expert observers}}.
\newblock {\emph{\JournalTitle{Psychological Science}}}
  \textbf{\bibinfo{volume}{24}}, \bibinfo{pages}{1848--1853},
  \url{10.1177/0956797613479386} (\bibinfo{year}{2013}).

\bibitem{bruno2015understanding}
\bibinfo{author}{Bruno, M.~A.}, \bibinfo{author}{Walker, E.~A.} \&
  \bibinfo{author}{Abujudeh, H.~H.}
\newblock \bibinfo{journal}{\bibinfo{title}{Understanding and confronting our
  mistakes: the epidemiology of error in radiology and strategies for error
  reduction}}.
\newblock {\emph{\JournalTitle{Radiographics}}} \textbf{\bibinfo{volume}{35}},
  \bibinfo{pages}{1668--1676} (\bibinfo{year}{2015}).

\bibitem{TobiiProSpark}
\bibinfo{author}{{Tobii Pro}}.
\newblock \bibinfo{title}{{Spark eye tracker} ({Pro Spark})}.
\newblock \bibinfo{howpublished}{{Apparatus and software}}
  (\bibinfo{year}{2019}).

\bibitem{dexl2026autopet}
\bibinfo{author}{Dexl, J.} \emph{et~al.}
\newblock \bibinfo{journal}{\bibinfo{title}{Autopet challenge on fully
  automated lesion segmentation in oncologic pet/ct imaging, part 2: Domain
  generalization}}.
\newblock {\emph{\JournalTitle{Journal of Nuclear Medicine}}}
  \textbf{\bibinfo{volume}{67}}, \bibinfo{pages}{481--488},
  \url{10.2967/jnumed.125.270260} (\bibinfo{year}{2026}).

\bibitem{hooge2024eye}
\bibinfo{author}{Hooge, I.~T.} \emph{et~al.}
\newblock \bibinfo{journal}{\bibinfo{title}{Eye tracker calibration: How well
  can humans refixate a target?}}
\newblock {\emph{\JournalTitle{Behavior Research Methods}}}
  \textbf{\bibinfo{volume}{57}}, \bibinfo{pages}{23} (\bibinfo{year}{2024}).

\bibitem{jacene2009assessment}
\bibinfo{author}{Jacene, H.~A.} \emph{et~al.}
\newblock \bibinfo{journal}{\bibinfo{title}{Assessment of interobserver
  reproducibility in quantitative 18f-fdg pet and ct measurements of tumor
  response to therapy}}.
\newblock {\emph{\JournalTitle{Journal of Nuclear Medicine}}}
  \textbf{\bibinfo{volume}{50}}, \bibinfo{pages}{1760--1769}
  (\bibinfo{year}{2009}).

\bibitem{constantino2023evaluation}
\bibinfo{author}{Constantino, C.~S.} \emph{et~al.}
\newblock \bibinfo{journal}{\bibinfo{title}{Evaluation of semiautomatic and
  deep learning--based fully automatic segmentation methods on [18f] fdg pet/ct
  images from patients with lymphoma: influence on tumor characterization}}.
\newblock {\emph{\JournalTitle{Journal of Digital Imaging}}}
  \textbf{\bibinfo{volume}{36}}, \bibinfo{pages}{1864--1876}
  (\bibinfo{year}{2023}).

\bibitem{kudura2021malignancy}
\bibinfo{author}{Kudura, K.} \emph{et~al.}
\newblock \bibinfo{journal}{\bibinfo{title}{Malignancy rate of indeterminate
  findings on fdg-pet/ct in cutaneous melanoma patients}}.
\newblock {\emph{\JournalTitle{Diagnostics}}} \textbf{\bibinfo{volume}{11}},
  \bibinfo{pages}{883} (\bibinfo{year}{2021}).

\bibitem{moses2011fundamental}
\bibinfo{author}{Moses, W.~W.}
\newblock \bibinfo{journal}{\bibinfo{title}{Fundamental limits of spatial
  resolution in pet}}.
\newblock {\emph{\JournalTitle{Nuclear Instruments and Methods in Physics
  Research Section A: Accelerators, Spectrometers, Detectors and Associated
  Equipment}}} \textbf{\bibinfo{volume}{648}}, \bibinfo{pages}{S236--S240}
  (\bibinfo{year}{2011}).

\bibitem{isensee2021nnunet}
\bibinfo{author}{Isensee, F.}, \bibinfo{author}{Jaeger, P.~F.},
  \bibinfo{author}{Kohl, S.~A.}, \bibinfo{author}{Petersen, J.} \&
  \bibinfo{author}{Maier-Hein, K.~H.}
\newblock \bibinfo{journal}{\bibinfo{title}{nnu-net: a self-configuring method
  for deep learning-based biomedical image segmentation}}.
\newblock {\emph{\JournalTitle{Nature methods}}} \textbf{\bibinfo{volume}{18}},
  \bibinfo{pages}{203--211}, \url{10.1038/s41592-020-01008-z}
  (\bibinfo{year}{2021}).

\bibitem{isensee2024nnunet}
\bibinfo{author}{Isensee, F.} \emph{et~al.}
\newblock \bibinfo{title}{nnu-net revisited: A call for rigorous validation in
  3d medical image segmentation}.
\newblock In \emph{\bibinfo{booktitle}{International Conference on Medical
  Image Computing and Computer-Assisted Intervention}},
  \bibinfo{pages}{488--498}, \url{10.1007/978-3-031-72114-4\_47}
  (\bibinfo{organization}{Springer}, \bibinfo{year}{2024}).

\bibitem{deng2009imagenet}
\bibinfo{author}{Deng, J.} \emph{et~al.}
\newblock \bibinfo{title}{Imagenet: A large-scale hierarchical image database}.
\newblock In \emph{\bibinfo{booktitle}{2009 IEEE Conference on Computer Vision
  and Pattern Recognition}}, \bibinfo{pages}{248--255}
  (\bibinfo{organization}{IEEE}, \bibinfo{year}{2009}).

\bibitem{dosovitskiy2021image}
\bibinfo{author}{Dosovitskiy, A.} \emph{et~al.}
\newblock \bibinfo{title}{An image is worth 16x16 words: Transformers for image
  recognition at scale}.
\newblock In \emph{\bibinfo{booktitle}{International Conference on Learning
  Representations}} (\bibinfo{year}{2021}).

\bibitem{vaswani2017attention}
\bibinfo{author}{Vaswani, A.} \emph{et~al.}
\newblock \bibinfo{journal}{\bibinfo{title}{Attention is all you need}}.
\newblock {\emph{\JournalTitle{Advances in neural information processing
  systems}}} \textbf{\bibinfo{volume}{30}}, \bibinfo{pages}{5998--6008}
  (\bibinfo{year}{2017}).

\bibitem{lin2017focal}
\bibinfo{author}{Lin, T.-Y.}, \bibinfo{author}{Goyal, P.},
  \bibinfo{author}{Girshick, R.}, \bibinfo{author}{He, K.} \&
  \bibinfo{author}{Doll{\'a}r, P.}
\newblock \bibinfo{title}{Focal loss for dense object detection}.
\newblock In \emph{\bibinfo{booktitle}{Proceedings of the IEEE international
  conference on computer vision}}, \bibinfo{pages}{2980--2988}
  (\bibinfo{year}{2017}).

\bibitem{isensee2025nninteractive}
\bibinfo{author}{Isensee, F.} \emph{et~al.}
\newblock \bibinfo{journal}{\bibinfo{title}{nninteractive: Redefining 3d
  promptable segmentation}}.
\newblock {\emph{\JournalTitle{arXiv preprint arXiv:2503.08373}}}
  (\bibinfo{year}{2025}).

\end{thebibliography}


\section*{Acknowledgements} 

\begin{itemize}
\item Nuclear Medicine and Molecular Imaging, Department of Radiology, Stanford University/Stanford Health Care.
\item Computer Vision and Machine Learning Systems, University of Münster.
\item Stanford AIMI Center for hosting the GazeXPErT dataset for open sourcing.
\item Google Academic Research Credits Program for data storage and compute resources in the cloud.
\end{itemize}

\section*{Author contributions statement}
All authors reviewed the manuscript. Specific contributions are as follow:
\begin{itemize}
\item Joy Wu (*first equal) - clinical concept conception, idea development and application framing for FDG-PET/CT, building data collection platform, sourcing hardware and team, collecting data, exploratory analysis, model development, literature review, writing and editing manuscript, coordinating and driving the project.
\item Daniel Beckmann (*first equal) - idea development, building data collection platform, main model development and training, exploratory analyses, literature review, writing and editing manuscript.
\item Sarah Miller - idea development, collecting data, coordinating project, literature review, writing and editing manuscript.
\item Alexander Lee - collecting data, independent data quality review, reviewing manuscript.
\item Elizabeth Theng - collecting data, exploratory data analysis, literature review, editing and reviewing manuscript.
\item Stephan Altmayer - collecting data, editing and reviewing manuscript.
\item Ken Chang - collecting data, reviewing manuscript.
\item David Kersting - collecting data, reviewing manuscript.
\item Tomoaki Otani - collecting data, reviewing manuscript.
\item Brittany Z Dashevsky - collecting data, reviewing manuscript.
\item Hye Lim Park - collecting data, reviewing manuscript.
\item Matteo Novello - collecting data, reviewing manuscript.
\item Kip Guja - collecting data, reviewing manuscript.
\item Curtis Langlotz - supervision, editing and reviewing manuscript.
\item Ismini Lourentzou - supervision, literature review, editing and reviewing manuscript.
\item Daniel Gruhl - idea development, key project development supervision, editing and reviewing manuscript.
\item Benjamin Risse (**Senior advisor) - idea development, main technical modeling supervisor, literature review, editing and reviewing manuscript.
\item Guido A Davidzon (**Senior advisor) - clinical concept conception, idea development and application framing for FDG-PET/CT, collecting data, project coordination, main clinical supervisor, literature review, editing and reviewing manuscript.
\end{itemize}

\section*{Competing interests} 
The authors declare no competing interests.

\section*{Figures \& Tables}
None additional.

\end{document}